\documentclass[]{aa}
\usepackage[utf8]{inputenc}
\usepackage{epsfig}

\usepackage{graphicx,natbib}
\usepackage{natbib,subfigure} 
\usepackage{amsmath,amssymb}
\usepackage[T1]{fontenc}

\usepackage{txfonts}

\usepackage{hyperref}

\hypersetup{
   colorlinks=true,
   citecolor=blue,
   linkcolor=blue,
   urlcolor=black
   }

\usepackage[english]{babel}

\def\Rgp{R}

\def\grav{g}

\def\Rp{R_\mathrm{p}}
\def\Rs{R_\star}

\def\X{X}
\def\Y{Y}
\def\Z{Z}
\def\Xobs{X_\mathrm{obs}}
\def\Yobs{Y_\mathrm{obs}}
\def\Zobs{Z_\mathrm{obs}}
\def\Xray{X_\mathrm{ray}}
\def\Yray{Y_\mathrm{ray}}
\def\Zray{Z_\mathrm{ray}}

\def\rc{\rho}
\def\th{\theta}
\def\x{x}
\def\xlimb{\x_\mathrm{limb}}
\def\z{z}
\def\zt{\z_\mathrm{t}}
\def\rs{r}
\def\lon{\lambda}
\def\lat{\varphi}
\def\colat{\alpha}
\def\lonobs{\lambda_\mathrm{obs}}
\def\lonstar{\lambda_{\star}}
\def\latobs{\varphi_\mathrm{obs}}
\def\colatobs{\alpha_\mathrm{obs}}

\def\vx{\hat{\mathbf{u}}_\mathrm{obs}}
\def\vX{\hat{\mathbf{X}}}
\def\vY{\hat{\mathbf{Y}}}
\def\vZ{\hat{\mathbf{Z}}}
\def\vR{\mathbf{u}}

\def\vray{\mathbf{u}_{\mathrm{ray}}}

\def\ix{i}
\def\Nrt{N_\x}
\def\dxrt{\Delta \x }
\def\dtaurt{\Delta \tau }
\def\taurt{\tau }
\def\trans{\mathcal{T}}
\def\Nc{N_\mathrm{con}}
\def\Ns{N_\mathrm{spe}}
\def\temp{T}
\def\tempi{T_i}
\def\press{P}
\def\pressi{P_i}
\def\pdeep{\press_\mathrm{bot}}
\def\ptop{\press_\mathrm{top}}
\def\chiji{\chi_{j,\ix}}
\def\chid{\chi_\mathrm{day}}
\def\chin{\chi_\mathrm{night}}
\def\M{M}
\def\qki{q_{k,\ix}}
\def\sigmol{\sigma_\mathrm{mol}}
\def\sigscat{\sigma_\mathrm{sca}}
\def\sigcont{\sigma_\mathrm{con}}
\def\kmie{k_\mathrm{mie}}
\def\sigmolj{\sigma_{\mathrm{mol},j}}
\def\sigscatj{\sigma_{\mathrm{sca},j}}
\def\sigcontj{\sigma_{\mathrm{con},j}}
\def\kmiek{k_{\mathrm{mie},k}}
\def\reff{r_\mathrm{eff}}
\def\Tday{\temp_\mathrm{day}}
\def\Tnight{\temp_\mathrm{night}}
\def\Tret{\temp_\mathrm{ret}}
\def\Trel{\Theta}
\def\H{H}
\def\dH{\widehat{\Delta \H}}
\def\dz{\widehat{\Delta \z}}\def\Hday{\H_\mathrm{day}}
\def\Hnight{\H_\mathrm{night}}
\def\Hdeep{\H_\mathrm{iso}}
\def\thstar{\alpha_{\star}}

\def\piso{\press_\mathrm{iso}}
\def\ziso{z_\mathrm{iso}}
\def\openAngle{\beta}
\def\limbAngle{\psi}
\def\tautr{\tau_\mathrm{tr}}
\def\tauvert{\tau_\mathrm{\perp}}

\def\ngasref{\ngas_0}
\def\ngas{n}

\def\sigobs{\sigma_\mathrm{obs}}
\def\sigobsl{\sigma_\mathrm{obs,\lambda}}
\def\d{\mathrm{d}}

\newcommand{\balign}[1]{
\begin{align}
#1
\end{align}}

\newcommand{\eq}[1]{Eq.\,(\ref{#1})}
\newcommand{\eqs}[2]{Eqs.\,(\ref{#1}) and (\ref{#2})}
\newcommand{\fig}[1]{Fig.\,\ref{#1}}
\newcommand{\figs}[2]{Figs.\,\ref{#1} and \ref{#2}}
\newcommand{\sect}[1]{Sect.\,\ref{#1}}

\newcommand{\app}[1]{Appendix\,\ref{#1}}
\newcommand{\tab}[1]{Table\,\ref{#1}}

\begin{document}

\title{
%Effects of the 3D structure of exo-atmopsheres on transmission spectra: Are we really/only probing the terminator?\\
%Biases in retrieval of atmospheric properties from transmission spectra due to day-night side temperature gradients \\
Effects of a fully 3D atmospheric structure on exoplanet transmission spectra: retrieval biases due to day-night temperature gradients}

\titlerunning{Effects of the 3D structure of exo-atmopsheres on transmission spectra}

\author{A. Caldas\inst{1}
     \and
           J. Leconte\inst{1}
           \and 
           F. Selsis\inst{1}
           \and
           I.P. Waldmann\inst{2}
           \and 
           P. Bordé\inst{1}
           \and
           M. Rocchetto \inst{2}
           \and
           B. Charnay \inst{3}}
           
\institute{Laboratoire d'astrophysique de Bordeaux, Univ. Bordeaux, CNRS, B18N, all\'{e}e Geoffroy Saint-Hilaire, 33615 Pessac, France
\and
Department of Physics and Astronomy, University College London (UCL), United Kingdom
\and
LESIA, Observatoire de Paris, PSL Research University, CNRS, Sorbonne Universit\'e, Univ. Paris Diderot, Sorbonne Paris Cit\'e, 92195, Meudon, France}
%\date{March 2017}

\abstract{
Transmission spectroscopy provides us with information on the atmospheric properties at the limb, which is often intuitively assumed to be a narrow annulus around the planet. Consequently, studies have focused on the effect of atmospheric horizontal heterogeneities \textit{along} the limb. Here we demonstrate that the region probed in transmission -- the limb -- actually extends significantly toward the day and night sides of the planet. We show that the strong day-night thermal and compositional gradients expected on synchronous exoplanets create sufficient heterogeneities \textit{across} the limb to result in important systematic effects on the spectrum and bias its interpretation. 
To quantify these effects, we developed a 3D radiative transfer model able to generate transmission spectra of atmospheres based on 3D atmospheric structures. We first apply this tool to a simulation of the atmosphere of GJ\,1214\,b to produce synthetic JWST observations and show that producing a spectrum using only atmospheric columns at the terminator results in errors greater than expected noise. This demonstrates the necessity of a real 3D approach to model data for such precise observatories. 
Second, we investigate how day-night temperature gradients cause a systematic bias in retrieval analysis performed with 1D forward models. For that purpose we synthesize a large set of forward spectra for prototypical HD\,209458\,b and GJ\,1214\,b type planets varying the temperatures of the day and night sides as well as the width of the transition region. We then perform typical retrieval analyses and compare the retrieved parameters to the \textit{ground truth} of the input model. This study reveals systematic biases on the retrieved temperature (found to be higher than the terminator temperature) and abundances. This is due to the fact that the hotter dayside is more extended vertically and screens the nightside---a result of the nonlinear properties of atmospheric transmission. These biases will be difficult to detect as the 1D profiles used in the retrieval procedure are found to provide an excellent match to the observed spectra based on standard fitting criteria. This fact needs to be kept in mind when interpreting current and future data. 
%In other words, in the presence of a strong day-night gradient, the region that we are effectively probing through transmission spectroscopy is not exactly at the terminator and can be significantly shifted toward the dayside. This fact needs to be kept in mind when interpreting current and future data. 
}


\maketitle

\section{Introduction}

\subsection{Biases in the analysis of transmission spectra of tridimensional planets}

With the first spectroscopic observations of exoplanets, we are now able to study planetary atmospheres beyond our solar system. In recent years, spectroscopic observations have seen tremendous developments, and the coming years are even more promising, particularly because of the launch of the James Webb Space Telescope (JWST; \citealt{BBK14}) and the dedicated ARIEL mission \citep{TDE17}.

One of the most common method to interpret atmospheric spectra is based on inverse atmospheric retrieval modelling \citep{Mad18}. However, because of the complex thermal structure of the atmosphere and the numerous gases to retrieve (with their possibly complex spatial distribution) the number of parameters to handle can render the inversion computational cost prohibitive---that is if there are enough data to have a well-constrained problem to start with. As a result, retrieval algorithm are required to make drastic assumptions on the forward model to render the problem tractable. 

\vspace{1cm}
The problem is that when these assumptions do not hold to a sufficient degree in the observed atmosphere, that creates a systematic bias that can lead the retrieval algorithm far from meaningful solutions. Identifying and alleviating these biases is thus a crucial goal to prepare for the next generation of precision observatories, and there have been several attempts in this direction. For example, the often made assumption of uniform mixing ratio in the atmosphere led \citet{ESK17} to retrieve a 100-1000$\times$ solar VO/H$_2$O ratio in the atmosphere of of WASP-121\,b. But \citet{PLB18} showed that accounting for the chemical dissociation of some species at the hottest altitudes allowed them to understand the data with solar abundances. \citet{RWV16} also thoroughly quantified the impact of assuming a vertically isothermal atmosphere.

Yet, all current retrieval algorithms are still fundamentally limited by the assumption of a spherically symmetric atmosphere: they use one-dimensional forward models to constraint spectra and atmospheric parameters of three-dimensional objects for which we expect heterogeneities. Such an approach is bound to create counter-intuitive biases that we need to quantify. With that in mind, \citet{FLF16} investigated how the non-uniform flux emitted by the planet could actually create a false positive signal for methane in emission. In the same vein, \citet{BDG17} used the dayside emission spectrum computed with a post-processed 3D Global Climate Model (GCM) to identify the region effectively probed by the retrieval of secondary eclipse data. 

On the transmission spectroscopy side, the study of the horizontal atmospheric heterogeneities have focused on the effect of clouds, with \citet{LP16} who have showed that the presence of clouds on parts of the limb only could mimic a high mean molecular weight atmosphere. Yet, this study produced their forward spectra by simply averaging two 1D models so that only a limited kind of heterogeneities could be investigated. To go further  \citet{CMM15}, \citet{WAA17}, \citet{PLB18}, and \citet{LMM18}, among others, have produced transit spectra from 3D atmospheric simulations. However, because of the difficult geometry, they still relied on a 1D radiative transfer transmission code that is either fed an average limb profile from a 3D simulation or that performs spectra of all the columns at the terminator of the model -- that they assume to be equivalent to the limb plane -- before averaging. Even if the second approach - that we will hereafter call \textit{limb-averaged} or (1+1)D method -- does capture the spatial variations of the atmosphere \textit{along} the terminator, it completely neglects horizontal variations \textit{across} it. Indeed, as the ray goes from the day side to the night side before coming to the observer, it crosses one of the most steeply changing region of the atmosphere: the transition from day to night side. The effect of such a thermo-compositional transition within the limb on retrieved parameters is unknown at present.
Indeed, although various authors have also developed a fully consistent transmission model able to predict such effects, these authors have not tried to retrieve physical parameters from their forward spectra. \citet{FSS10}, \citet{BRS10}, and \citet{DAB12}, for example, have focused on the potential differences between the east and west limbs, while \citet{MR12} and \citet{SFL13} have mainly looked at the effect of the doppler shifting by winds on high resolution spectra.

\subsection{How wide can the limb be: a simple estimate}

\begin{figure}
   \centering
   \includegraphics[scale=.3,trim = 6cm .cm 4.5cm 11.cm, clip]{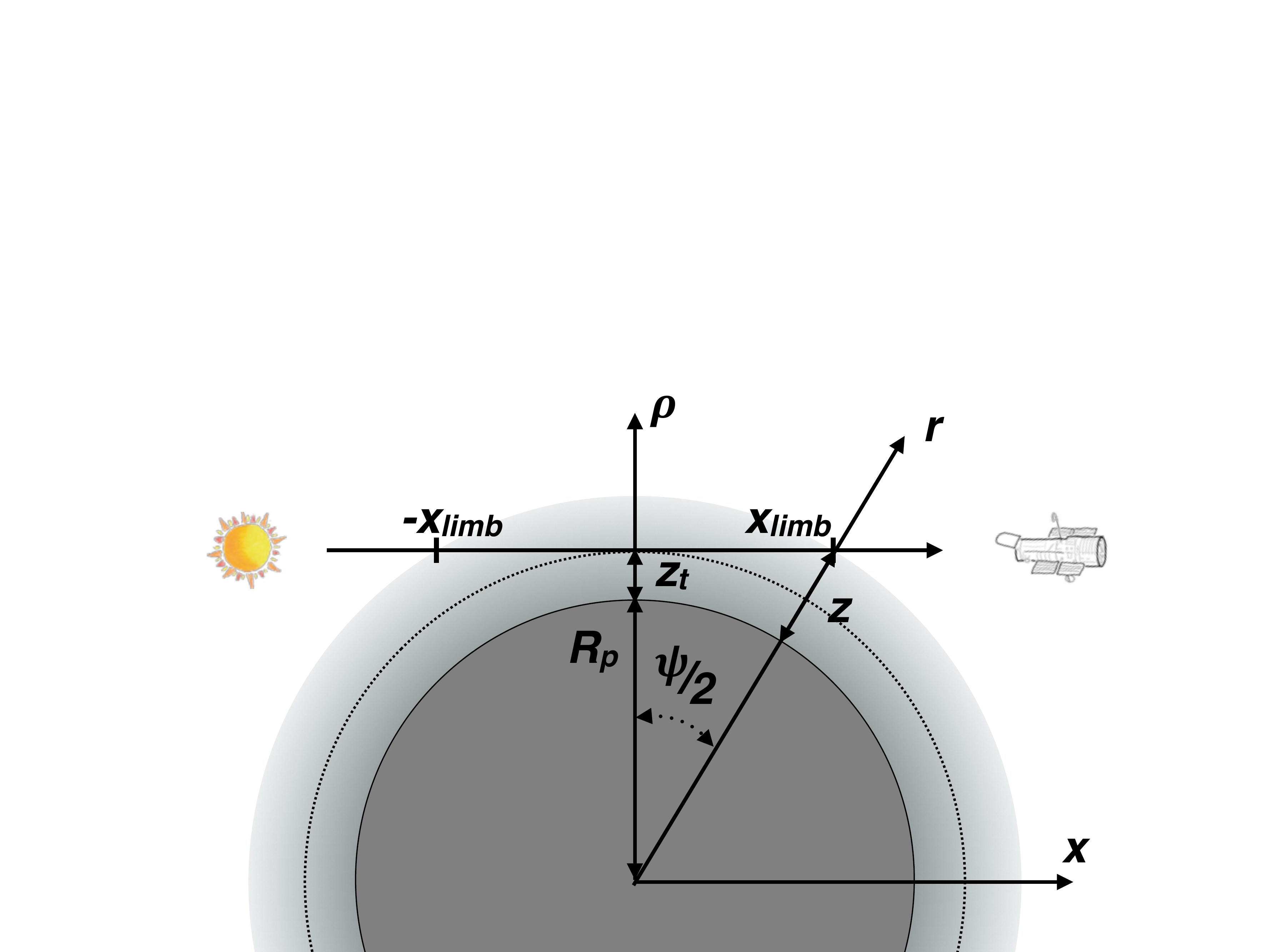}
      \caption{
      Schematic of the geometry of a light ray crossing the atmosphere. The inner circle is the arbitrary reference surface of the planet of radius $\Rp$. The lighter grey region is the atmosphere. The distance from the center of the planet is $\rs=\Rp+z$. A light ray is defined by its distance of closest approach to the planet's center, $\rc$, and the corresponding tangential altitude $\zt=\rc-\Rp$. The direction of the light ray defines the $\x$ direction. As we further discuss the extent of the limb along the ray, we introduce $\xlimb$ so that the absorption outside the $[-\xlimb,\xlimb]$ segment is negligible in determining the transit radius, and the corresponding limb opening angle $\limbAngle$.
      }
         \label{schematic_transit}
   \end{figure}

The day-night transition would not be a problem if the region probed in transmission were infinitely thin. We would just see a slice of the atmosphere. 
But in fact, and quite counter-intuitively, the width of this region -- that will be our definition of the limb\footnote{The limb, as defined here, should not be confused with either i) the limb plane which is the plane perpendicular to the observer's line of sight passing through the planet center or ii) the terminator that is also a plane passing through the planet center, but which is perpendicular to the star-planet axis. The latter two are confounded only when the star, planet, and observer are aligned. } -- is much larger on some planets that is generally expected. Therefore, the transit spectrum encodes a much wider diversity of temperatures and compositions. 

Although the effect of this larger extent of the limb will be demonstrated \textit{a posteriori} by the results of our 3D transit model, let us here try to give simple arguments to estimate how different planets can be affected. In other words, how wide can we expect the limb to be on any given planet. 

Of course, the problem in providing such a simple estimate is that the region that will contribute to the transit spectrum does not only depend on the global parameters of the planet, but also on the precise chemical-physical conditions in the atmosphere and how they vary spatially, as will be demonstrated later on. There is thus a certain degree of arbitrariness if one wants to come up with a simple general estimate.
For this reason, we will first use a simple geometrical argument. The advantage of this is that it will allow us to identify the key dimensionless parameter controlling the limb width. In \app{app:guillotmodel} we derive a model of a more specific case of chemical inhomogeneity and show that the two approaches indeed yield similar results.

Let us consider a light ray passing through the limb as shown in \fig{schematic_transit}. Estimating the width of the limb comes down to the computation of the maximum distance from the limb plane, $\xlimb$, at which the atmosphere still affects \textit{measurably} the optical depth along all the observed rays, and especially the deepest one. The choice we have to make here is the highest pressure probed in transit ($\pdeep$, that we will assume to define the planetary radius $\Rp$) and the lowest pressure at which the atmosphere is still able to significantly affect the transmission of a given ray ($\ptop$). Then the width of the limb is given by
\balign{\limbAngle\equiv 2 \arccos \left(\frac{\Rp}{\Rp+\z(\ptop)}\right).}
Since, in a isothermal atmosphere with a atmospheric scale height at the surface equal to $\H$ and a varying gravity, the hypsometric relation writes
\balign{
\z(\ptop)-\z(\pdeep)=\H \ln \left(\frac{\pdeep}{\ptop}\right)\left(\frac{1}{1-\frac{\H}{\Rp}\ln\left(\frac{\pdeep}{\ptop}\right)}\right),
}
we get
\balign{\limbAngle\equiv 2 \arccos \left(1-\frac{\H}{\Rp}\ln \left(\frac{\pdeep}{\ptop}\right)\right).\label{geometricallimb}}

The first important result is that we see that the important dimensionless quantity in our problem is the ratio of the scale height to the planetary radius. The higher this parameter, the larger the curvature effects in the atmosphere. This parameter will often appear later on.
We also directly see that the larger the pressure range probed, the wider the limb. Many models predict that the lowest levels probed in transit are around 100\,mb in the visible/near infrared. On the other hand, \citet{KBD14} have shown that in order to explain the flat spectrum of GJ\,1214\,b, an opaque aerosol deck is needed as high as 10$^{-3}$-10$^{-2}$\,mb, showing that absorbers at such altitudes can indeed still affect the transit spectrum. 

%%%%%%%%%%%%%%%%%%%%%%%%%%%%%%%%%%%%%%%%
%%%%%%%%%%%%%%%%%%%%%%%%%%%%%%%%%%%%%%%%
\begin{figure}[hbtp] %  figure placement: here, top, bottom, or page
 \centering
\subfigure{ \includegraphics[scale=.75,trim = 0cm .cm 0.cm 0.cm, clip]{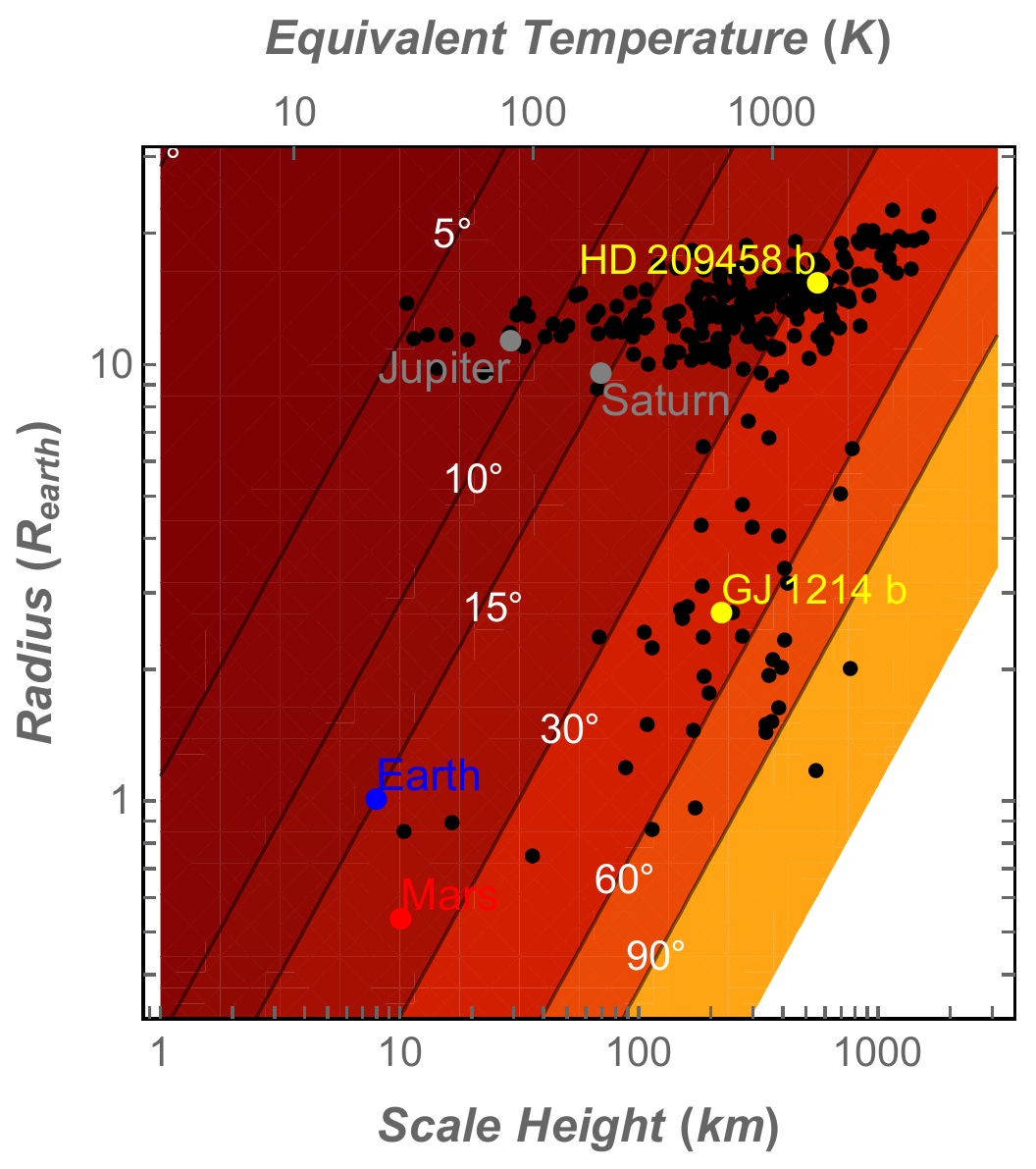} }
\caption{
Simple estimate of the opening angle of the region around the terminator that affects the transmission spectrum (i.e. the limb). The contours give the opening angle in degrees as a function of the planetary radius and the atmospheric scale height of the atmosphere at 10\,mb (the variation of gravity with height is accounted for). The atmosphere is assumed to become transparent above the 1\,Pa pressure level. Black dots are known planets for which the radius and surface gravity are measured. Some key objects are identified. The equivalent temperature is derived from the scale height assuming an atmosphere of hydrogen and helium (solar abundances) and a surface gravity of 10\,m/s$^2$. For hot, low gravity planets like GJ\,1214\,b, the limb can cover almost half the surface of the planet!}
 \label{Cells}
\end{figure}
%%%%%%%%%%%%%%%%%%%%%%%%%%%%%%%%%%%%%%%%
%%%%%%%%%%%%%%%%%%%%%%%%%%%%%%%%%%%%%%%%

To get a quantitative idea, \fig{Cells} shows the result of \eq{geometricallimb} as a function of the planetary radius and scale height of the atmosphere. To have a relatively conservative estimate, we restrained the pressure range to ($\pdeep,\,\ptop$)=(10\,mb,\,10$^{-2}$\,mb).
One can see that even for the Earth, the limb region already spans more than ten degrees. For a warm sub-Neptune planet like GJ\,1214\,b, this can be as large as 45$^\circ$-50$^\circ$. In numerical terms, this means that within a typical resolution 3D atmosphere model of this planet, say 128 grid points in longitude for sake of concreteness, a single ray would interact with about 24 consecutive horizontal cells before leaving the planet. From these numbers, it becomes evident that the 1+1D approach, by picking only one out of those 24 cells as representative of the terminator, is a crude approximation to a real transmission spectrum.

\subsection{Goals of the study}

Our goal is thus to identify the various biases of retrieval methods created by thermal and compositional -- including clouds -- inhomogeneities in the atmosphere in transmission. To that purpose, we need a transmission spectrum generator able to match the complexity of a real three-dimensional planet. We thus developed a tool able to compute transmission spectra using a parametrized 3D atmospheric structure or the outputs of a 3D atmospheric simulation by a global climate model---namely the LMD Generic model \citep{WFS11,LFC13b,CMM15}. This tool, Pytmosph3R, and its architecture are described in  \sect{Presentation}. Then we show an extensive validation in \sect{sec:validation}.
Next, \sect{Application} presents a first application of this tool to a simulation of the atmosphere of GJ\,1214\,b where we demonstrate the necessity of a real 3D approach to model data for such precise observatories. 
Finally, we investigate how day-night temperature gradients expected for exoplanets cause a systematic bias in retrieval analysis of real data performed with 1D forward models (\sect{daynightretrieval}).

\section{Presentation of Pytmosph3R}\label{Presentation}

Pytmosph3R is designed to compute transmission spectra based on 3D atmospheric simulations performed with the LMDZ generic global climate model. It produces transmittance maps of the atmospheric limb at all wavelength that can then be spatially integrated to yield the transmission spectrum.  
The code is entirely written in python.

In this section we present the various modules of the code:
\begin{itemize}
    \item[$\bullet$] The geometrical framework used to map the atmospheric structure from the spherical coordinates used by the GCM onto cylindrical coordinates that are more suitable to follow photons crossing the atmosphere,
    \item[$\bullet$] The two algorithms that can be used for the calculation of the slant optical path -- a discretised and an integral method,
    \item[$\bullet$] The various sources of opacity included in our radiative transfer model,
    \item[$\bullet$] The spatial integration to produce spectra.
\end{itemize}

\subsection{Definition of coordinate systems}

\subsubsection{The spherical grid used in atmospheric simulations}\label{sphere}

Typical 3D atmospheric simulations -- LMDZ included -- provide state variables such as temperature and mixing ratios of various absorbers/scatterers on a longitude/latitude/pressure grid. Although pressure is convenient variable to compute atmospheric motions, transit spectroscopy is fundamentally about knowing the physical area of the opaque region of the atmosphere. We thus first have to interpolate the outputs of the climate model on a spherical ($\lon,\,\lat,\,z$) grid, where $\lon$ is the longitude, $\lat$ is the latitude, and $z$ the altitude. When needed, we will also refer to $\rs\equiv \Rp+z$ the distance to the planet center, or $\colat$ the colatitude. 

The longitude/latitude grid is evenly spaced and follows the native grid of the atmospheric model. The altitude grid is also evenly spaced. However, as will be discussed in detail in \sect{sec:validation}, the resolution of this grid can, and usually should, be higher than the native resolution of the input simulation as it will set the precision of the output spectrum. We find that a good compromise between computation time and accuracy is reached for a vertical resolution of about a tenth of the scale height. Let $n_z$, $n_{\lon}$, and $n_{\lat}$ be the number of grid cells in each of the three dimensions. 

The top should also be chosen high enough for the atmosphere to be transparent there. This will be quantified hereafter. If this altitude is above the top of the input simulation, the code extrapolates the atmosphere above this top assuming hydrostatic equilibrium and a fixed temperature (or a profile that the user needs to define).

Then, we integrate the hydrostatic equilibrium equation within each column of the model to compute the values of all the necessary variables from the climate model (e.g. temperature and mixing ratios) on this new altitude grid. During this integration, the variation of gravity with altitude is taken into account (see \eq{hypsometric}), an effect that proved to be crucial to reach the precision needed.

\subsubsection{The cylindrical grid used for the radiative transfer}\label{cyl}

As long as the light from the star propagates in straight lines, cylindrical coordinates provide the most natural set to treat transit geometry. Indeed, due to the great distance between the observer and the planet, light rays follow parallel paths so that the radiative transfer can be solved within a hollow cylinder that is tangent to the planetary surface at the bottom and stops at the top of the modelled atmosphere.

\begin{figure*}
   \centering
   \includegraphics[width=18cm]{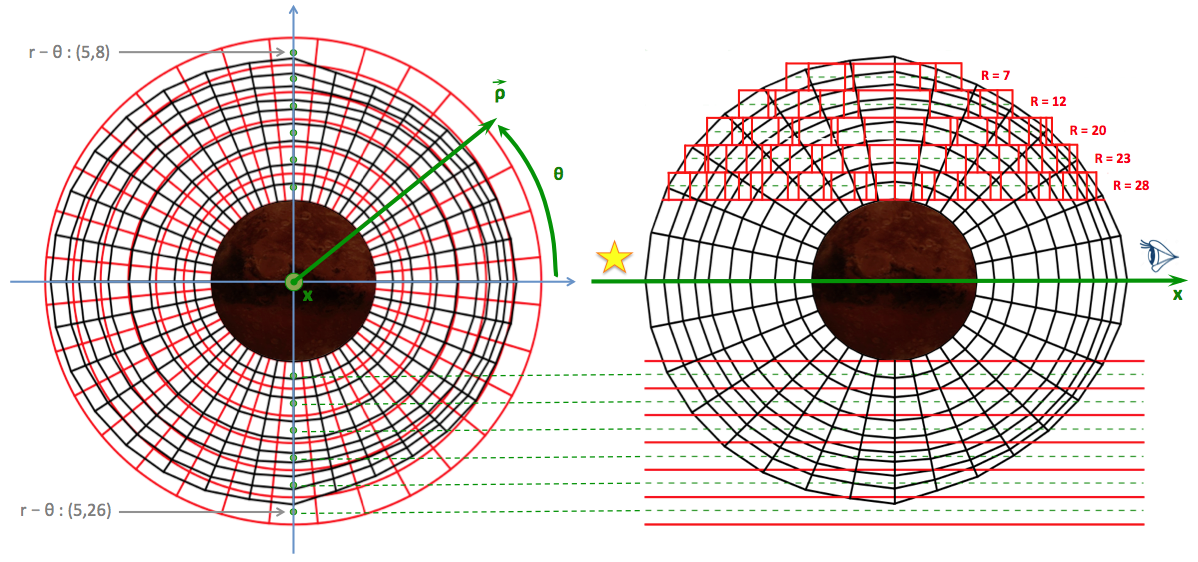}
      \caption{Schematic showing the cylindrical grid (in green). The left panel shows the grid as seen from the observer. The planet moves to the right, which defines the $\th=0$ axis. The pink dots are examples of ($\rc,\,\th$)-rays (To follow the python convention used by the code, the indices given start at 0). The native GCM grid at the terminator of an example simulation is shown in black to illustrate its non-uniform vertical spacing. The right panel shows this from the side where we can follow light rays (dashed lines) as they go horizontally from left to right. The red boxes illustrate how the $x$-discretisation is computed for each rays to follow closely where it goes from on cell of the spherical grid to another. 
              }
         \label{Proj}
   \end{figure*}

We thus define a cylindrical grid whose coordinates will be denoted ($\rc,\,\th,\,x$) where the $x$ axis is the line between the center of the planet and the observer (the line of sight), with $x$ increasing toward the observer (see \fig{Proj}). The $x$=0 plane corresponds to the plane perpendicular to the line of sight passing through the center of the planet (hereafter called the plane of the sky).
$\rc$ is the distance from the $x$ axis in the plane of the sky, and $\th$ is the azimuth, defined here as the angle on the limb plane measured from a reference direction (see below).

Hereafter we will refer to a "($\rc,\,\th$)-ray" as the ray of light that crosses the limb plane at those coordinates. 
The transmittance map of the atmosphere is simply constructed by computing the chord optical depth of the ray along the $x$ direction for every ($\rc,\,\th$).
 From this, and assuming a luminosity/spectral distribution (limb-darkening, spots) over the stellar disk as well as a transit trajectory, we can produce spectral transit lightcurves.

%\textbf{(Franck: maybe it would be better to start with the observing geometry itself. Are there some limitations in terms of obliquity/inclination and position on the stellar disk?) CHECK}

%%%%%%%%%%%%%%%%%%%%%%%%%%%%%%%%%%%%%%%%
%%%%%%%% Figure 2
%%%%%%%%%%%%%%%%%%%%%%%%%%%%%%%%%%%%%%%%
%%%%%%%%%%%%%%%%%%%%%%%%%%%%%%%%%%%%%%%%
%%%%%%%% Figure 2
%%%%%%%%%%%%%%%%%%%%%%%%%%%%%%%%%%%%%%%%

 The resolution of the cylindrical grid is based on that of the spherical one, which has a finer altitude resolution compared to the GCM: $\Delta \rho$=$\Delta z$ and $\Delta \theta$=$\Delta \varphi$. There is no benefit in increasing the resolution in $\theta$, GCM cells being considered uniform horizontally. To test the impact of shooting our rays through the middle of our layers or at their interfaces, the $\rho$ grid can be shifted relative to the $r$ grid using the $\omega$ parameter ($0 \leq \omega < 1$) so that $\rho = r +\omega\,\Delta z$. To speed up computation, $(\rho,\theta)$-rays will eventually be divided irregularly along the $x$ direction to follow closely where they go from one spherical cell to another, as explained in \ref{subpaths} (see also \fig{Proj}).

\subsubsection{Orientation of the planet and correspondence between coordinate systems}\label{orientation}

The  cylindrical coordinate system needs to be properly oriented with respect to the spherical grid. For this, we require only to know the longitude and latitude of the observer in the spherical coordinates, $(\lonobs,\latobs)$, at the time of observation (or alternatively the colatitude $\colatobs$)\footnote{In the simple case of a planet with a null obliquity, $\latobs=| \pi /2-i|$, where $i$ is the orbital inclination with respect to plane of the sky. The sign depends on the convention for $i$. Furthermore, at conjunction, $\lonobs=\lonstar+\pi$, where $\lonstar$ is the substellar longitude.  }. The unit vector pointing toward the observer, $\vx$, then defines the direction of the $x$ axis of our cylindrical coordinates.

The last thing that we need to define is the arbitrary reference direction for the azimuth. For this we choose the projection of the rotation axis of the planet onto the plane of the sky. 

With these two definitions, there is a unique relationship between the spherical and cylindrical coordinates for a given point. The translation from one system to the other with an arbitrary orientation however requires to solves a set of non-linear equations that is detailed in \app{app:coordsystem}.

\subsection{Dividing $(\rho,\theta)$-rays into subpaths}\label{subpaths}
%Opacity calculations are performed along $(\rho,\theta)$-rays parallel to $x$ and intercepting the geometrical center of each cell in the $\rho - \theta$ plane. 
Along each $(\rc,\th)$-ray (identified by the indices $i_{\rc}$ and $i_{\th}$) we locate all the intersections with relevant interfaces of the spherical-grid and divide the ray into segments of irregular lengths, each of them belonging to a different cell. All quantities used to calculate the optical path -- pressure and density excepted -- are considered constant within each segment/cell. The pressure/density can be either kept constant within a segment/cell or assumed to follow hydrostatic equilibrium.

In practice we first divide each $(\rc,\th)$-ray with a constant $\Delta x$ step, calculate the spherical coordinates of the resulting discrete points and give them the three indices ($i_\rs$,$i_\lambda$,$i_\varphi$) of the spherical cell they belong to (see \app{app:coordsystem}). The step $\Delta x$ is chosen small enough ($< \Delta z$) so that two successive points can only belong to either the same cell or to two adjacent cells (i.e., cells separated by a facet, an edge, or a corner). One, two, or all three indices ($i_\rs$,$i_\lon$,$i_\lat$) can be incremented between two successive points. 

When a change of index occurs between two points, the code determines analytically the position of the intersection(s) between the $(\rc,\th)$-ray and the surface(s) separating cells. This comes down to solving for the unknown position $\x_\mathrm{int}$ along the ray knowing $\rc$, $\th$, and the equation of the surface(s) crossed. The equations to be solved for the three type of intersections (depending on the varying index) are detailed in \app{app:intersections}

From these positions, the length of the subpaths belonging to individual cells can be measured. When more than one index is incremented between two points (near an edge or a corner, which implies that a third cell has been crossed) and once the intersections have been located, their x-position are sorted in increasing order so that subpaths can be measured and attributed to specific cells.

\subsection{Optical depth}

At this point, all the $(\rc,\th)$-rays are now subdivided into $\Nrt(\rc,\th)$ segments of length $\{\dxrt_\ix (\rc,\th)\}_{\ix=1,\Nrt}$. The number and length of these segments of course changes for each $(\rho,\theta)$-rays depending on the number of intersections found in the previous step. 
Each of these segments has been assigned to a given cell of the spherical grid so that we know all the quantities describing the physical state of the atmosphere in the $\ix$-th segment: temperature ($\tempi$), volume mixing ratio of the $j$-th of the $\Ns$ species ($\chiji$), and mass mixing ratio of the $k$-th of the $\Nc$ species of condensed particles ($\qki$). 

The goal is now to compute the optical depth (hence the transmittance) of the atmosphere for each $(\rc,\th)$ which is given by
\balign{\taurt (\rc,\th)=\sum_{\ix}^{\Nrt}\dtaurt_{\ix}(\rc,\th),}
where $\dtaurt_{\ix}$ is the optical depth of a given segment. 

Pytmosph3R can calculate optical depth in two ways. Pressure (and density) can be either considered constant within a cell, hereafter called the \textit{discretised} method, or it can be integrated along the optical path assuming a hydrostatic vertical structure within the cell, the so-called \textit{semi-integral} method. In both cases, we assume that cross-sections are constant along the given sub-path to spare considerable CPU time. This approximation yields negligible errors as long as the simulation is sufficiently resolved in the vertical.

\subsubsection{Discretized calculation of the optical depth}

With the \textit{discretized} method, pressure and density are considered constant so that the optical depth of a segment simply reduces to
\balign{\label{discret}
\dtaurt_\ix=\left(\frac{\pressi}{k_B \tempi}\sum_j^{\Ns}\chiji \, (\sigmolj+\sigscatj+\sigcontj)+\sum_k^{\Nc} \kmiek\right)\dxrt_\ix
}
where $\sigmol$, $\sigscat$, and $\sigcont$ are the cross-sections for the molecular, Rayleigh scattering, and continuum absorptions. $\kmie$ is the absorption coefficient associated to the Mie scattering by aerosols. The parametrization used for these absorptions is discussed hereafter. 

\subsubsection{Integral method}

With the \textit{integral} method, we now assume that the pressure follows the hydrostatic law within a segment, thus varying with altitude. The optical depth of a segment for any contribution is then given by
\balign{\label{integral}
\dtaurt_\ix&=\int_{\x_\ix}^{\x_\ix +\dxrt_\ix}\frac{\press(\x)}{k_B \tempi} \,\chi \,\sigma \,\mathrm{d} \x 
=\frac{\chi \,\sigma}{k_B \tempi} \, \int_{\x_\ix}^{\x_\ix +\dxrt_\ix}\press(\x) \mathrm{d} \x ,
}
where $\x_\ix$ is the positions of the beginning of the segment. Let us call $z_\ix$ the corresponding altitude. The simplification comes from the fact that the altitude profile of the pressure within an isothermal atmosphere is analytical even if the gravity varies with height. It is given by
\balign{\label{hypsometric}
\ln \left(\frac{\press(z_{i+1})}{\press(z_i)}\right) = \int_{z_i}^{z_{i+1}}\frac{-\M \grav(z_i)}{\Rgp \temp}\frac{\mathrm{d}z}{\left(1+\frac{z-z_i}{\Rp+z_i}\right)^2} = -\frac{\M \grav(z_i)}{\Rgp \temp}\left(\frac{z_{i+1}-z_i}{1+\frac{z_{i+1}-z_i}{\Rp+z_i}}\right),}
where $\M$ is the molar mass of the gas, $\Rgp$ the universal gas constant, and $\grav(z)$ the gravity at a given altitude. This entails
\balign{
\int_{\x_\ix}^{\x_\ix +\dxrt_\ix}\press(z(\x)) \,\mathrm{d} \x  &= \int_{\x_\ix}^{\x_\ix +\dxrt_\ix}\press(z_1)\exp\left[-\frac{\M \grav(z_\ix)}{\Rgp \tempi}\left(\frac{z-z_\ix}{1+\frac{z-z_\ix}{\Rp+z_\ix}}\right)\right]\mathrm{d} \x \nonumber\\
&= \int_{z_\ix}^{z_\ix +\Delta z_\ix}\press(z_\ix)\exp\left[-\frac{\M \grav(z_\ix)}{\Rgp \tempi}\left(\frac{z-z_\ix}{1+\frac{z-z_\ix}{\Rp+z_\ix}}\right)\right]\times\nonumber\\
&\hspace{3cm} \times\frac{(\Rp+z)}{\sqrt{(\Rp+z)^2-\rho^2}}\,\mathrm{d} z,}
%\balign{\frac{1}{k_BT(\delta,\phi,h)}\sum_i^{N_s}\chi_{i,\delta,\phi,h}\sigma_{m,i,\delta,\phi,h}\int_{z_1}^{z_2}P(h_1)exp\left(\frac{h-h_1}{H(h)}\right)dz,}
where the last integration is carried out numerically between the lowest and the highest point of the segment. The increased accuracy of this method comes from two reasons: i) the variation of gravity with height is built in, and ii) more importantly the  exponential variation of pressure is fully accounted for. This explains why a much lower number of vertical layers are needed with this method to reach numerical convergence.    
%Taking into account of the altitude z dependence with the gravity and after a small change of variable
%\balign{&\frac{1}{H(z)}=\frac{Mg_{z_1}}{RT}\times\left(\frac{1}{1+\frac{z-z_1}{\Rp+z_1}}\right)^2\\
%&\tau_{\rho,\theta,x}(z,\lambda,\varphi)=\nonumber\\
%&\frac{P_{z_1}}{k_BT}\sum\sigma
%\int_{z_1}^{z_2}\exp\left[-\frac{Mg_{z_1}}{RT}\left(\frac{z-z_1}{1+\frac{z-z_1}{\Rp+z_1}}\right)\right]\frac{(\Rp+z)dz}{\sqrt{(\Rp+z)^2-\rho^2}},}

\subsection{Sources of opacity}

\subsubsection{Molecular lines}

  Pytmosph3R deals with molecular absorptions in two possible ways: tables of monochromatic cross-sections or correlated-$k$ coefficients (also called $k$-distribution method; \citealt{FU}).
 
 Using $k$-distributions considerably reduces the computing time but new $k$-tables must be pre-computed each time one wants to change the atmospheric composition or the resolution of the output spectrum. 
 
 Because our code is designed to work with the TauREx retrieval code \citep{wald}, we use the same set of high-resolution cross-sections produced by the ExoMol project. This data set is precomputed on a $T-\log P$ grid going from 200\,K to 2800\,K every 100\,K and from $10^{-3}$ to 10\,bar every 0.3\,dex. Pythmosph3R users can either choose linear or \textit{optimal} interpolation in temperature. For the optimal interpolation we follow \citet{Hill} who prescribes
\balign{\label{sigmalambda}
\sigma_{i,\lambda}(T)=a_{i,\lambda} \exp\left(-\frac{b_{i,\lambda}}{T}\right),
}
where $i$ is the molecular/atomic species index, $\lambda$ the wavelength, and $T$ the temperature of the cell. The $(a,b)$ scaling factors are given by  
\balign{
&b_{i,\lambda}=\left(\frac{1}{T_l}-\frac{1}{T_u}\right)^{-1}ln\frac{\sigma_{i,\lambda}(T_u)}{\sigma_{i,\lambda}(T_l)}\label{moneq}\\
&a_{i,\lambda}=a_{i,\lambda}(T_u)\exp\left(\frac{b_{i,\lambda}}{T_u}\right),}
where $T_u$ and $T_l$ are upper and lower temperatures respectively. $k$-distributions can only be interpolated linearly in temperature. Along the pressure coordinate, the interpolation is log-linear.

The effect of using different interpolations schemes has been tested and we find that it does not introduce significant differences.

\subsubsection{Continuum absorptions}

Our principal source of continuum opacity is due to collision induced absorptions. We account for this process for all the species for which such information is available in the HITRAN database and following the prescriptions of \citet{RGR11}. Furthermore, for specific species such as water vapor, we can add a continuum that is accounting for the truncation of the far wings of the lines and the neglect of many weak lines in some of our sets of cross-sections or correlated-$k$ tables. In such case, the water continuum is added using the CKD model \citep{CKD89}. We however note that care must be taken to ensure that the molecular opacities used must be computed consistently to do not count some effects twice. 

\subsubsection{Rayleigh Scattering}

Multiple scattering is neglected. The contribution of Rayleigh scattering is thus treated as a simple extinction. The cross-section of any single gas molecule is given by the common formula
\balign{\sigscat(\lambda)=\frac{24\pi^3 }{N_\mathrm{std}^2 \lambda^4 }\left(\frac{n_\lambda^2-1}{n_\lambda^2+2}\right)^2F_k(\lambda),}
where $\lambda$ is the wavelength (here in $m$), $N_\mathrm{std}$ is the number density of a gas under standard conditions, $n_\lambda$ is wavelength-dependent, real refractive index of the gas and $F_k(\lambda)$ is the King correction factor which accounts for the depolarization. The accuracy of this essential part of the radiative transfer mainly depends of the calculation of refraction indices. We used the most recent data available in the literature. For sake of completeness, we have reviewed the parametrization that we use for H$_2$, He, H$_2$O, N$_2$, CO, CO$_2$, CH$_4$, O$_2$ and Ar in \app{rayleighscatdata} and in \tab{index}.

\subsubsection{Mie scattering for aerosols}

Transmission spectra of transiting exoplanets are affected by clouds and hazes. The LMDZ GCM can include cloud physics and provide the properties and 3D distribution of liquid/solid condensates and aerosols. Assuming spherical particles with a size similar to or larger than the considered wavelengths, we use Mie scattering formalism compute their extinction factor $Q_{ext}$ and resulting opacities, following the same method as in radiative transfer modules of the GCM \citep{MFM11}. We linearly interpolate the value $Q_{ext}$ on effective radius and wavelength using pre-calculated lookup tables. The absorption coefficient is estimated as
\balign{\kmiek(\lambda)=\frac{3}{4}\frac{Q_\mathrm{ext}(\reff,\lambda)}{\rho_\mathrm{con} \reff} q_{k} \,\rho_\mathrm{gas},}
where $Q_{ext}$ is the extinction coefficient for the wavelength $\lambda$ and a given effective radius $\reff$, $q_{k}$ and $\rho_\mathrm{con}$ are the mass mixing ratio and the density of the species considered, and $\rho_\mathrm{gas}$ the total gas density.

\subsection{Generation of transmittance maps and spectra}\label{observables}

To generate the global absorption spectrum of the planet, the wavelength-dependent $(\rc,\th)$ map of optical depth is first converted into a transmittance map
\balign{\trans(\rc,\th,\lambda)=e^{- \taurt (\rc,\th,\lambda)}.}
In the most general case, the in-transit flux should be computed by convolving this transmittance map with a given surface brightness distribution for the star.
However, in the most simple case of a homogeneous stellar disk the effective area of the planet reads 
\balign{A_p(\lambda)&=\pi \Rp^2+ \sum_{\rc,\th}\left(1-\trans(\rc,\th,\lambda)\right)\,S_{\rc}\nonumber\\
&=\pi \Rp^2+ \sum_{\rc,\th}\left(1-e^{-\taurt(\rc,\th,\lambda)}\right)S_{\rc},}
where $\Rp$ is the radius of planet (at the bottom of our model atmosphere), $S_{\rc}=2\pi (\rc+\Delta \rc/2)\Delta \rc/N_{\th}$, $N_{\th}$ the number of $\th$ points, and $\Delta \rho$ the layer thickness. Eventually, the relative dimming of the stellar flux due to the planet is given by
\balign{\label{Reff}
\Delta F(\lambda)=\frac{A_p}{\pi \Rs^2}=\left(\frac{\Rp}{\Rs}\right)^2+\sum_{\rc,\th}\frac{\left(1-e^{-\taurt(\rc,\th,\lambda)}\right)S_{\rc}}{\pi \Rs^2}}
where $\Rs$ is the stellar radius. 
For each monochromatic transit depth, we can determine an effective radius of the planet defined as 
\balign{
R_\mathrm{eff}(\lambda) = \Rs\sqrt{\Delta F(\lambda)}.
}

One can simulate a realistic light curve by locating precisely the planet transmittance map over the weighted stellar disk at each timestep of the transit. During ingress and egress, where the stellar disk is hidden by a fraction only of the planet, only the elements in front of the stellar surface need to be considered. Then, astrophysical (stellar) noise can be added by simulating a realistic spatial distribution of the star luminosity and its variability during a transit, e.g. by taking into account limb darkening, spots, and granulation \citep{CCS17}.
This model complements other codes able to produce synthetic emission/reflection spectra and light curves from GCM simulations \citep{Selsis2011,Turbet2016,Turbet2017}.
%%%%%%%%%%%%%%%%%%%%%%%%%%%%%%%%%%%%%%%%
%%%%%%%% Figure 4
%%%%%%%%%%%%%%%%%%%%%%%%%%%%%%%%%%%%%%%%
\begin{figure*}
   %\centering
   \sidecaption
   \includegraphics[width=12cm]{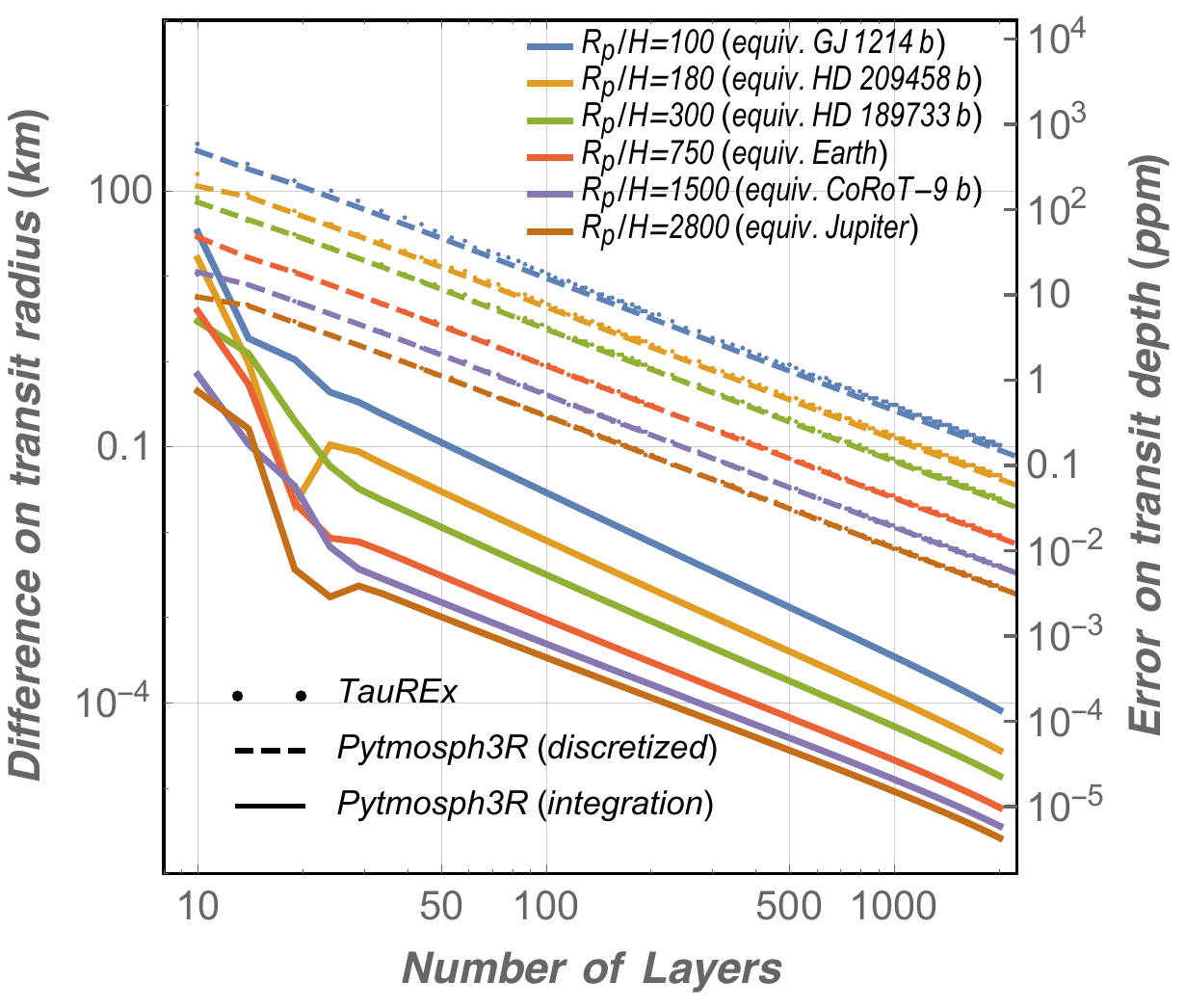}
      \caption{Absolute difference between the transit radius computed for a given number of layers and model and our reference case (Integral method, isothermal, grey opacity, 50\,000 layers). The dots show the results of TauREx, the dashed lines represent Pytmosph3R in its \textit{discretized} version, and the solid lines are for the \textit{integral} version.
      Results are plotted for 6 values of $\Rp/\H$ that increase from top to bottom for each set of curves. The names of planets representative of these values are labelled.
      For Pytmosph3R, we use the results of one sector only ($\theta = 0^\circ$) to keep the computing time reasonable for the most resolved simulations. The axis on the right converts this radius difference into a transit depth difference in the case of GJ\,1214\,b  (i.e. a 2.6 earth radius planet in front of a 0.2 solar radius star). To rescale this for any type of planet or star, multiply by $(\Rp/2.7 R_E)(0.2 R_\odot /\Rs)^2$. 
              }
         \label{Conv}
   \end{figure*}
%%%%%%%%%%%%%%%%%%%%%%%%%%%%%%%%%%%%%%%%
%%%%%%%% Figure 4
%%%%%%%%%%%%%%%%%%%%%%%%%%%%%%%%%%%%%%%%

\section{Model validation}\label{sec:validation}

\subsection{Validation approach}

In the following subsections we validate the different modules of the code: calculation of the vertical structure, 3D geometry of the radiative transfer, integration scheme for the optical depth, interpolation scheme for the opacities, etc. In the absence of another validated code able to produce synthetic transmission spectra from 3D simulations, we decided to compare our code with a hierarchy of 1D models with increasing complexity---from a purely analytical model (\sect{sec:validationguillot}) to the more realistic forward spectrum generator used in the TauREx retrieval tool (\citealt{wald}; \sect{sec:taurex}). 

In each case, the comparison with a given 1D model proceeds as follows: On one hand we generate a spectrum from the 1D model with a given vertical temperature profile (possibly isothermal). On the other hand, we use the same vertical profile to generate a full spherically symmetric-3D structure on our spherical grid. This structure goes through our whole code to generate a spectrum that should, ideally, be the same as the 1D one. Finally, we increase the resolution of our 1D/3D grids until convergence is reached.

It appeared relatively early in our tests that, to numerical precision, our transmittance map were completely insensitive to the azimuth angle, $\th$, in any spherically symmetric configuration, as it should be. This not completely trivial result shows that our careful way of computing the path length of the ray in each cell does get rid of the singularities -- in particular at the poles -- of our initial longitude/latitude grid. This also allowed us to test the convergence of the algorithm of vertical integration at very high resolutions.
Indeed, when a very large number of layers is used -- above $\sim$1000, which is already much larger than what needs to be used in practice to be converged -- the computing time for the full 3D code becomes prohibitive. For this reason, in some cases below, the results shown are derived from a sector of the limb only (i.e. a given $\theta$). This approach is made possible by the fact that the numerical differences between the various sectors is negligible.

As expected from our analytical arguments and further demonstrated in the next subsections, we find that the most, and in fact almost only, important parameter determining the resolution needed to model a given atmosphere is the ratio of the planetary radius to the scale height ($\Rp /\H$): the lower this parameter, the larger the vertical resolution needed. It stems from two reasons. A lower ratio means that i) gravity will vary more significantly in each vertical layer, and ii) curvature effects will be more pronounced and rays will pass through more atmospheric columns along their path (see \fig{Cells}). As a result the validation cases we present hereafter focus on covering a wide variety of $\Rp /\H$. However, because the scale height is not set, but determined from the composition and temperature of the atmosphere and from the surface gravity, fixing $\Rp /\H$ entails varying one of these parameters to compensate. For each of the validation setups described in the following sections, the compensation parameter is thus always stated. This can sometimes result in a physically inconsistent set of planetary parameters that we deem acceptable in the context of our validation. The values used for the other parameters are given in \tab{Param}, unless otherwise stated. 

As will become clear in the following, our 3D model can reproduce our validation cases to numerical precision. We thus used these tests to further derive some guidelines on the number of layers and the model roof pressure to use in various cases to reach a satisfactory accuracy. To quantify what we call a "satisfactory accuracy", we use the difference on the predicted transit depth between two models and ask that it be significantly smaller than the photon noise that can be expected for the given target in one transit. If one wants a more stringent condition, one can use the expected systematic noise floor for a given instrument (expected to be on the order of a few ppm with JWST, for example).

\subsection{Validation with a monochromatic, analytical model at constant gravity} \label{sec:validationguillot}

To focus on the radiative transfer basics we first compare the optical depths computed with our model with those given by an analytical solution. For an isothermal,  horizontally uniform atmosphere with constant gravity and a scale height $\H \ll \Rp$, \citet{Guigui} provides an analytical formula for the transmission optical depth as a function of the distance to the center of the planet: 
%altitude. This study will allow us to validate the writing of the radiative transfer and the pressure profile generator. Indeed, we do not integrate the optical paths following the same philosophy as 1D models because of the equations that are numerically adapted to the three-dimensional character of the atmospheric parameters. Let us admit now that the atmosphere is composed of a single active specie, supposing that its absorption is independent from pressure and temperature. As part of these assumptions, the effective radius of this exoplanet is reckoned as
\balign{
\tautr=\tauvert\left(\frac{2\pi \Rp}{\H}\right)^{\frac{1}{2}},}
where $\tautr$ is the chord optical depth, $\tauvert$ the vertical optical depth. 
%This expression gives us a first-order approximation to the effective radius valid in the limit $\H/\Rp\ll 1$.
As described in \app{app:guillotmodel}, we use this formula to calculate the planet transit. We then compare the results with those of Pytmosph3R for a wide range of vertical resolution: 50 to more than 10\,000 levels. For 1000 levels and more, and as long as the thin atmosphere assumption is respected, the effective absorption radii from both models agree within $\sim 1$\,centimetre (i.e. $\sim$ 10$^{-9}$ relative accuracy). %Therefore, we choose atmospheric parameters in order to comply with this condition, in other words a small, with a small scale height. \\

%\textbf{Franck: Why don't we simply compare the transmissivity as a function of $\rho$? As this is what Guillot et al. provide, why do we use the additional calculation of the effective radius that reduces the comparison to a single number while we can compare $T(\rho)$ profiles. I suggest to : (1) compare $T(\rho$) obtained from Guillot and from Pytmosphr3R for different values of $H0/\Rp$ and at very high spatial resolution so we can see a perfect agreement at low $\H/\Rp$ and an increasing departure towards higher $\H/\Rp$ due to the Guillot approximation. (2) to show the effect of spatial resolution, which can be done as you already did using only $R_eff$ or using $T(\rho)$. }
%We can perform a similar calculation with Pytmosph3R assuming one single 1D column covering the whole planet, with the same physical properties. 

%%%%%%%%%%%%%%%%%%%%%%%%%%%%%%%%%%%%%%%%
%%%%%%%% Table 2
%%%%%%%%%%%%%%%%%%%%%%%%%%%%%%%%%%%%%%%%
\begin{table}[htbp]
\centering
\caption{Numerical values for the parameters used in our fiducial validation case. }
\begin{tabular}{lll}
\hline \hline
Planetary Radius & $\Rp$ [m] & $7\times10^7$\\
Surface gravity & $g_0$ [m$^2$/s]& $8.8$ \\
Surface pressure & $P_s$ [Pa]& 1$\times10^{6}$\\
Top pressure & $P_{top}$ [Pa]& 1$\times10^{-4}$\\
He mass fraction && 0.17\\
Water volume mixing ratio &$\chi_{H_2O}$& 0.05\\
Mean molar mass & $M_{a}$ [kg/mol]& 2.7$\times 10^{-3}$\\
Opacity of gray absorber &$\kappa$ [m$^2$.kg$^{-1}$]&$1\times 10^{-3}$ \\
\hline \hline\\
\end{tabular}\label{Param}
\end{table}
%%%%%%%%%%%%%%%%%%%%%%%%%%%%%%%%%%%%%%%%
%%%%%%%% Table 2
%%%%%%%%%%%%%%%%%%%%%%%%%%%%%%%%%%%%%%%%

\subsection{Comparison to the TauREx forward model}\label{sec:taurex}

\subsubsection{Why benchmarking against the forward model of an atmospheric retrieval tool}

As one of our main goals is to be able to perform a retrieval on a spectrum derived from a 3D model as if it were from a real planet, we decided to use the forward model of an existing, and validated, atmospheric retrieval code for our comparison.

The added benefit of this validation approach is that we make sure that in the relevant case of spherically symmetric atmosphere, both our 1D and 3D forward model can produce spectra without any numerical biases---meaning here that the differences between the spectra can always be reduced to be much smaller than the noise that will be prescribed in the retrieval step by choosing a sufficient vertical resolution. 

As a result, we know that when we retrieve the properties of an atmosphere generated with our 3D model, the various potential biases in the retrieved parameters are entirely due to the heterogeneities of the atmosphere and not the differences in the numerics. 

In our case, we are using TauREx, whose forward model is described in more details in \citet{wald}. For the reasons mentioned above, we decided to use the same philosophy used in TauREx in the implementation of many physical processes, in particular concerning the molecular opacities and the Rayleigh scattering. 
Two notable exceptions are i) the vertical grid and ii) the optical path calculations that will be discussed more specifically in the next section. Let us just mention here that our requirement that our 3D model be as compatible with TauREx as possible is the reason why we keep two algorithms for the calculations of the optical depth in the code: the \textit{integral} scheme which is the optimal scheme in terms of convergence and should be used in general, and the \textit{discretised} one that always gives a result closer to TauREx when the same number of layers are used in both models.

%%%%%%%%%%%%%%%%%%%%%%%%%%%%%%%%%%%%%%%%
%%%%%%%% Figure 6
%%%%%%%%%%%%%%%%%%%%%%%%%%%%%%%%%%%%%%%%
\begin{figure*}
   \sidecaption%\centering
   \includegraphics[width=12.5cm]{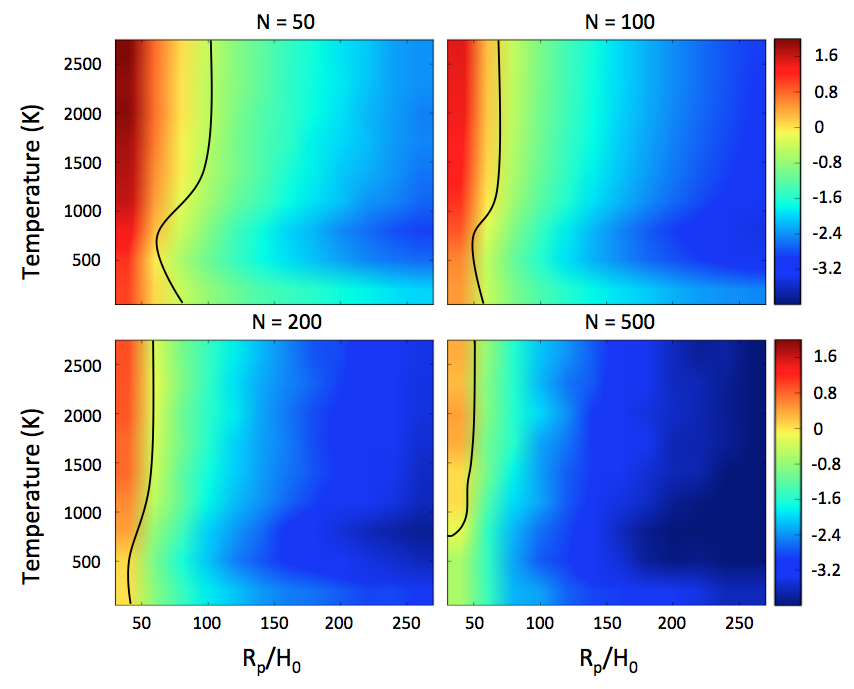}
      \caption{The figure shows the decimal logarithm of the reduced chi-square for the effective radius between our 3D model and the full forward model from TauREx over a range of temperature between $200$ and $2600$\,K and of $\Rp/\H$ between $40$ and $260$. As a comparison, all planets described in \fig{Conv} have $\Rp/\H>100$. The black line shows where the reduced $\chi^2$ equals unity, i.e. where model differences become insignificant compared to the noise. We assumed an error calculated with a sun-like star at 100 lyr, 1 hour of integration and JWST instruments. The same work has been done for four vertical resolutions : 50, 100, 200 and 500 layers. 
              }
         \label{Valid1}
   \end{figure*}
%%%%%%%%%%%%%%%%%%%%%%%%%%%%%%%%%%%%%%%%
%%%%%%%% Figure 6
%%%%%%%%%%%%%%%%%%%%%%%%%%%%%%%%%%%%%%%%

\subsubsection{Validation for a isothermal, gray atmosphere with varying gravity}

We now assume an isothermal atmosphere with an altitude-dependent gravity. We consider a uniform composition of H$_2$/He/H$_2$O. In this first step, to validate only the geometrical part of the code, we assume that H$_2$ and He are transparent, and that water has a gray opacity that is independent of temperature and pressure. 
In this set of simulations, the scale height is set to the prescribed value by adjusting the temperature of the atmosphere, all other parameters being kept fixed to the value given in \tab{Param}.

For each $\Rp/\H$ ratios, we first compute the transit radius of a high resolution reference model using Pytmosph3R with 50\,000 layers. Then we compute the difference between this reference and the transit radius given by our three models at various vertical resolutions. This is summarized in \fig{Conv} where the results for Pytmosph3R/discretised method are shown by dashed lines, for Pytmosph3R/integral method by solid lines, and TauREx by dots. The various colors are for different $\Rp/\H$. Note that although we quote some planet names for each ratio to give an idea of what type of planet it describes, calculations where all performed for the same planetary radius (1\,$R_J$).

The conclusions that can be drawn from this test are the following:
\begin{itemize}
    \item[$\bullet$]All three models converge toward the same result as vertical resolution is increased so that the differences can be reduced to an arbitrarily low value,
    \item[$\bullet$]As advertised, atmospheres with a low $\Rp/\H$ require less vertical layers to reach a given accuracy,
    \item[$\bullet$] For the same number of layers, the accuracy of our integral version of the code is orders of magnitude better than the discretized one. This is thus the preferred mode for most applications,
    \item[$\bullet$] As expected, the discretised mode, although less accurate, is always closer to TauREx and should probably be used if a retrieval analysis is to be performed. This indeed introduces as little biases as possible during the retrieval step.
\end{itemize}

In terms of the number of layers needed to reach convergence -- the precision needed depending of course on the observations -- the integral mode of Pytmoshp3R requires as little as 20-30 layers in almost all practical cases of interest. This yields about 2-3 points per pressure decade or about 1 per scale height. For the discretised version as well a for TauREx and indeed most forward model using the same philosophy, 100 layers are generally enough, but this should be taken with caution. Especially hot or low gravity objects require a finer resolution and some published models have probably reached convergence only marginally. 

Despite our efforts, it can be seen that for a given number of layers, the discrete version of Pytmosph3R is not exactly equivalent to TauREx. We find that this small discrepancy is fundamentally due the fact that our vertical grid uses altitude rather than pressure levels as usually done in 1D models. When gravity varies with altitude, an iso-altitude grid is not equivalent to an iso-log pressure one, hence the small difference.  

%These results are obviously modulated by the ratio $\Rp/\Rs$. For a given precision, 0.1 ppm for example, we will need more layers if we study a bigger exoplanets or systems with smaller host stars, and inversely.

\subsubsection{Comparison with the full TauREx forward model}

%%%%%%%%%%%%%%%%%%%%%%%%%%%%%%%%%%%%%%%%
%%%%%%%%%%%%%%%%%%%%%%%%%%%%%%%%%%%%%%%%
\begin{figure}
   \centering
   \includegraphics[width=9cm]{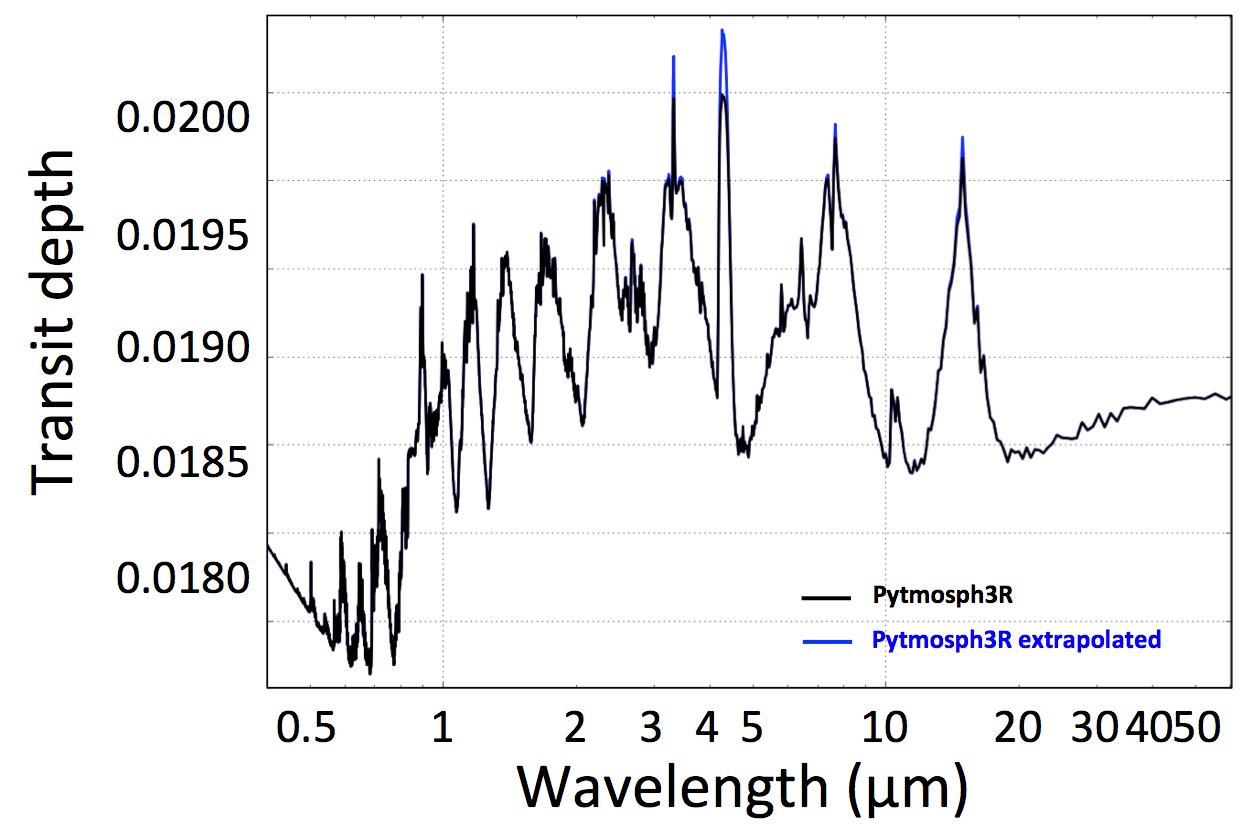}
      \caption{Transit spectrum generated by Pytmosph3R from the outputs of a GCM simulation of GJ\,1214\,b \citep{CMM15}. The black curve represents the effective radius obtained when accounting only for the part of the atmosphere explicitly modelled by the GCM (Down to $\sim$ 0.5\,Pa). The blue curve shows the result when the model is extended by an isothermal atmosphere down to 10$^{-4}$\,Pa. Further lowering the pressure of the model top does not alter the spectrum. }
         \label{Spec2}
   \end{figure}
%%%%%%%%%%%%%%%%%%%%%%%%%%%%%%%%%%%%%%%%
%%%%%%%%%%%%%%%%%%%%%%%%%%%%%%%%%%%%%%%%
Here we compare spectra from Pytmosph3R and TauREx for isothermal atmospheres with the same composition as before. There are two differences however. First, we now compute a full spectrum with all our opacity sources varying with wavelength, temperature, and pressure. Second, we now fix the atmosphere temperature beforehand and adapt the surface gravity of the planet to get the desired $\Rp/\H$ ratio. This stems from the fact that will now impact the molecular features directly and not only through the scale height. This allows us to test our interpolations in the opacity databases as well. 

As the model produces a full spectrum and not only a monochromatic transit radius, we quantify the differences between the two codes using a  reduced chi-square calculated as follows: 
\balign{
\chi^2=\sum_{\lambda}^{N_{\lambda}}\frac{(\delta_{3D,\lambda}-\delta_{1D,\lambda})^2}{\sigobsl^2}\frac{1}{N_{\lambda}}
}
where $\delta_{3D,\lambda}$ and $\delta_{1D,\lambda}$ are the transit depths at the wavelength $\lambda$ given by Pythmosph3R and TauREx, $N_{\lambda}$ is the number of spectral bins, and $\sigma_{h\nu,\lambda}$ is the stellar photon noise computed for a JWST observations. The latter is given by
\balign{
\sigobsl &=\frac{1}{\sqrt{N_{\mathrm{obs},\lambda}}},}
where
\balign{
N_{\mathrm{obs},\lambda} &= \frac{\pi\tau_{\lambda}c\Delta tR_{s}^2A}{2d^2}\int_{\lambda_{wl}}^{\lambda_{wl+1}} \frac{d\lambda}{\lambda^4\left(\exp(\frac{h_Pc}{k_BT_s\lambda}-1)\right)}.
}
The parameters are $\tau_{\lambda}$ is the system throughput, $\Rs$ and $T_s$ the radius and temperature of the host star, $A$ the collecting area of the telescope (here 25 m$^2$), $\Delta t$ the integration time, and $d$ the distance of the target. In our example, we considered a Sun-like star, a 1-hour integration, and a system at 100\,ly. Increasing the noise tends towards making Pytmosph3R and  TauREx spectra indistinguishable. 

We computed the logarithm of this reduced chi-squared as a function of the atmospheric temperature and $\Rp/\H$ for four different vertical resolutions. Results are shown in \fig{Valid1}.

As already discussed, we can see that the agreement between the two codes at a given vertical resolution generally increases with $\Rp/\H$. At a given $\Rp/\H$ ratio and in this specific example, the agreement also slightly depends on temperature. This is due to the overall increase of H$_2$O opacity with temperature that pushes the opaque region upward and increases the limb opening angle (see\fig{Cells}). With high enough vertical resolution the two models can however agree well within the noise budget. As the lowest encountered $\Rp/\H$ is about 100 (for a hot-Neptune like GJ\,1214\,b), we confirm that 50 layers should be sufficient in most cases. These maps can however be used as a guide to chose the resolution needed if a more stringent case is found. 

\subsection{Effect of model top altitude}

When computing transmission spectra, if the model does not reach a height in the atmosphere that is transparent enough at all wavelengths, the transit radius of the planet may be underestimated in opaques parts of the spectrum. The choice of the model top pressure thus generally results from a trade-off  between computation time and convergence.  

This point is even more crucial here because several other technical reasons can limit the maximum altitude of a 3D climate model to a few tenths of pascals or more (short radiative timescales, absence of non-LTE radiative treatment or conductive heat transfer, etc.).

To illustrate these limitations, \fig{Spec2} shows two spectra computed by Pytmosph3R for a simulation of GJ\,1214\,b (see \sect{Application}). The black spectrum uses directly the outputs of the global climate model and stop around a pressure level of 0.5\,Pa. The blue spectrum is obtained by extending the model upward to 10$^{-4}$\, Pa assuming that the atmosphere remains isothermal above the GCM model top. For this particular case, this is necessary and sufficient to reach a degree of convergence commensurate with the future precision of the data ($\sim 1-10$ppm). Extending the model to even lower pressures does not significantly change the resulting spectrum. 
\begin{figure}[tb] %  figure placement: here, top, bottom, or page
 \centering
\subfigure{ \includegraphics[scale=.7,trim = 2cm .cm .cm 0.cm, clip]{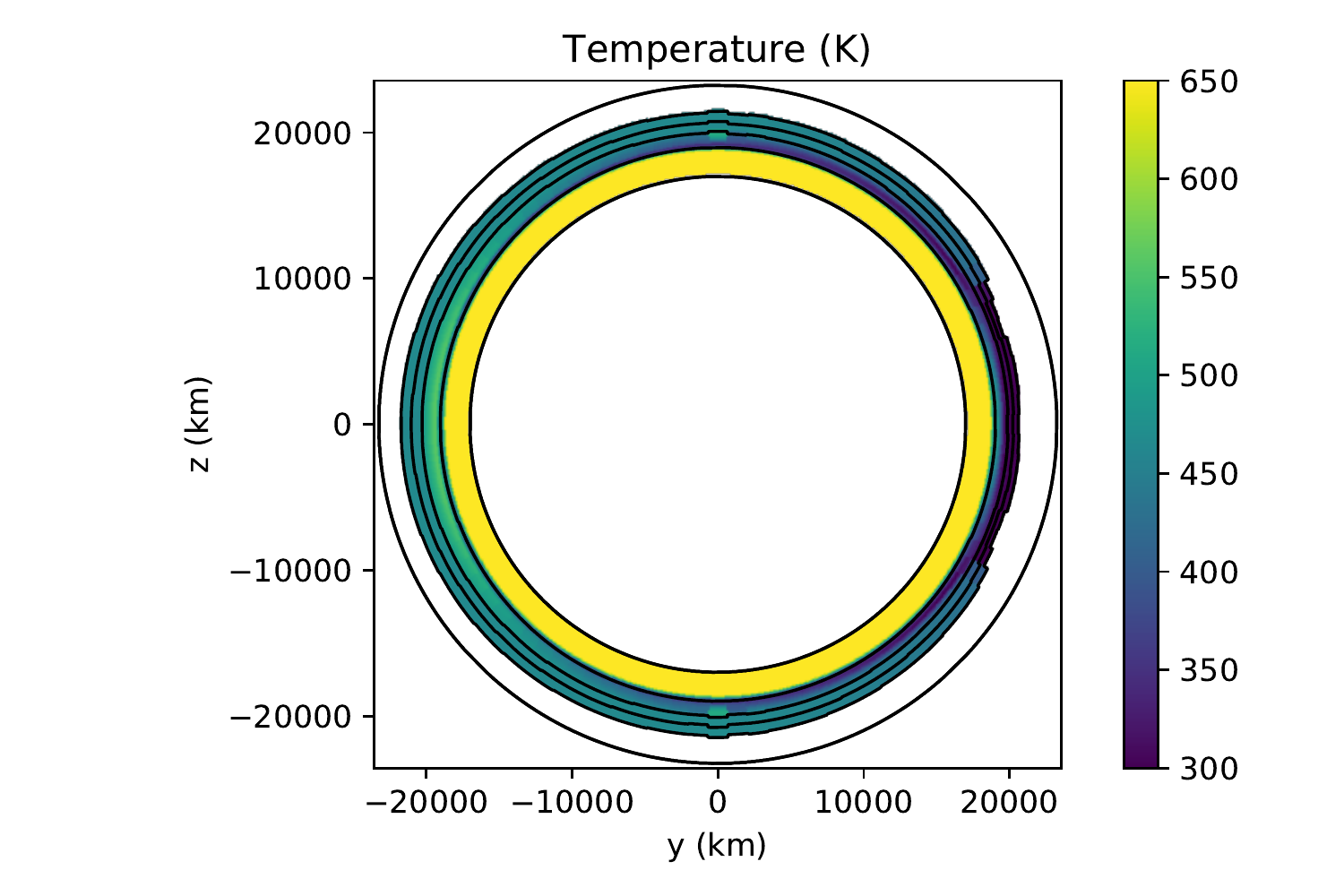} 
}
\subfigure{ \includegraphics[scale=.7,trim = 2.cm .cm 0.cm .75cm, clip]{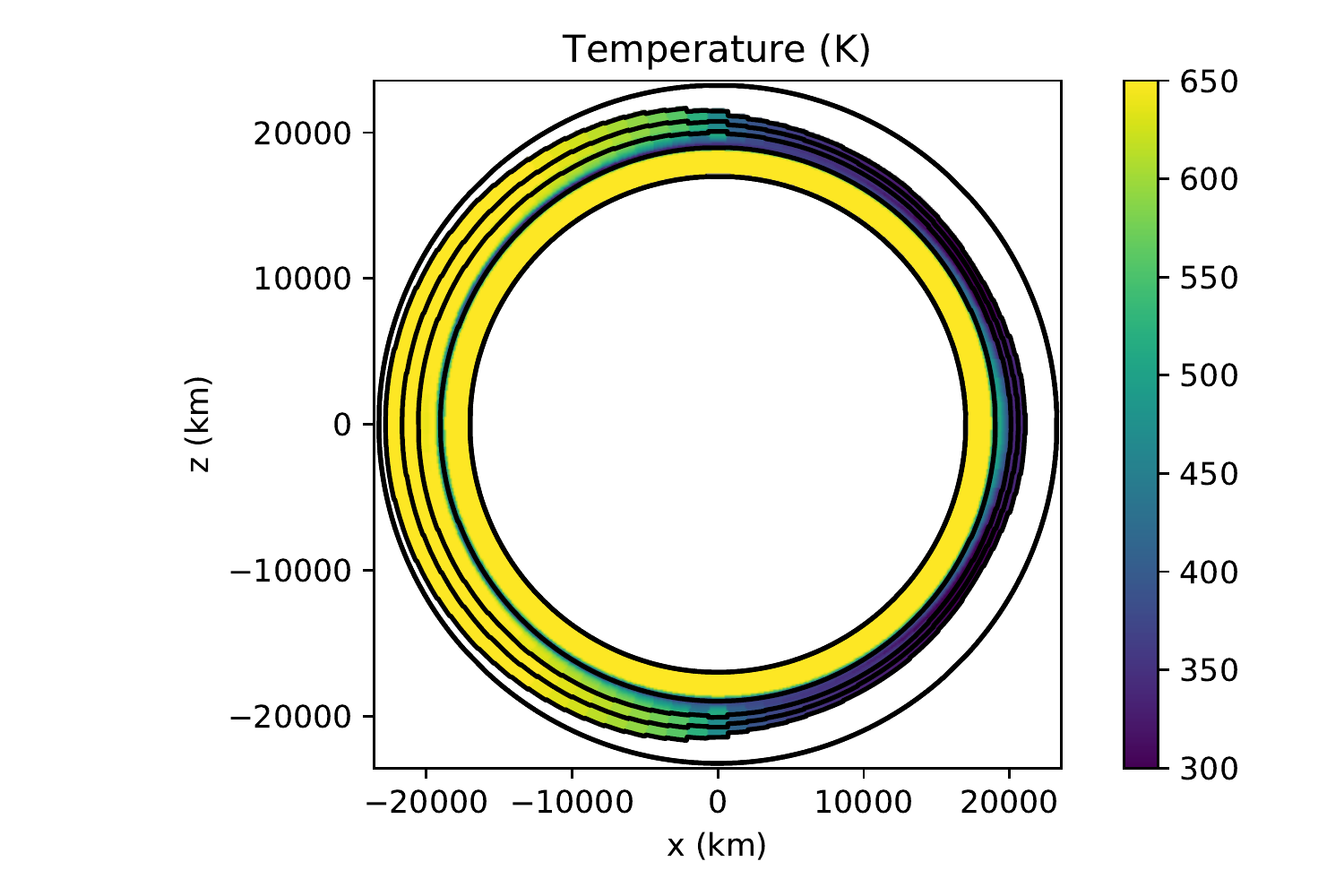} }
\caption{
Colour maps showing the temperature distribution in the model atmosphere of the 100\,$\times$ solar metallicity simulation from \citet{CMM15}. The inner white disk represents the inner part of the planet with a radius assumed to be equal to the one of GJ\,1214\,b (17.600\,km). Top: Temperature at the terminator. The planet is seen from the observer during transit, the poles being at the top and bottom. Bottom: The star is on the left and the observer on the right on the $z=0$ axis. The poles are on the $x=0$ axis. From center outward, the 5 solid lines are respectively the 10$^6$, 10$^3$, 1, 10$^{-2}$, and 10$^{-4}$\,Pa pressure levels. The outer circle is there as an eye guide to highlight the non-sphericity of the planet. Temperature is well homogenized below $\sim$10$^3$\,Pa level. Maps are \textit{to scale} and show that the dayside is noticeably more extended vertically than the nightside.  }
 \label{Tmap_GJ}
\end{figure}

%%%%%%%%%%%%%%%%%%%%%%%%%%%%%%%%%%%%%%%%
%%%%%%%%%%%%%%%%%%%%%%%%%%%%%%%%%%%%%%%%
\begin{figure*}[h]
   \centering
   
     \includegraphics[width=1.\linewidth]{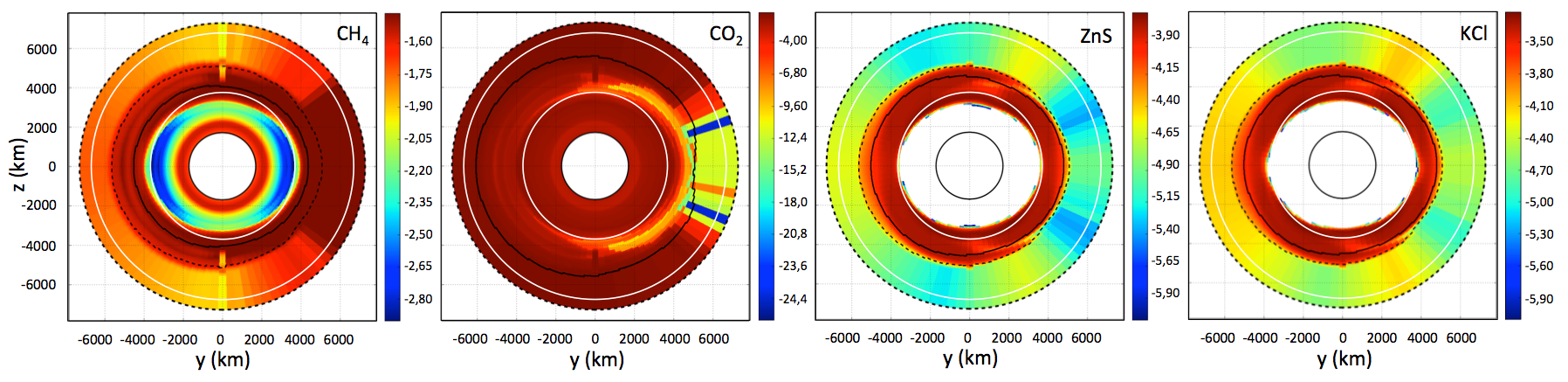}
      \caption{Distribution of the logarithm of the mass mixing ratio of some absorbing species at the terminator in a simulation of the atmosphere of GJ\,1214\,b by \citet{CMM15}. For KCl and ZnS, only the molecules in the condensed phase are considered. The dashed (solid) black curves are the contour of unit optical depth with (without) cloud opacities at a wavelength relevant to the molecule considered (1.17\,$\mu$m for CH$_4$; 4.30\,$\mu$m for CO$_2$; 1.08\,$\mu$m for KCl et ZnS). The size of the inner white disk delimiting the base of the atmosphere is reduced for clarity. The whole simulated atmosphere is shown. The two white circles delimit the region enlarged in \fig{maps1214}. }
         \label{fig.abs}
    \vspace{0.5cm}

     \includegraphics[width=1.\linewidth]{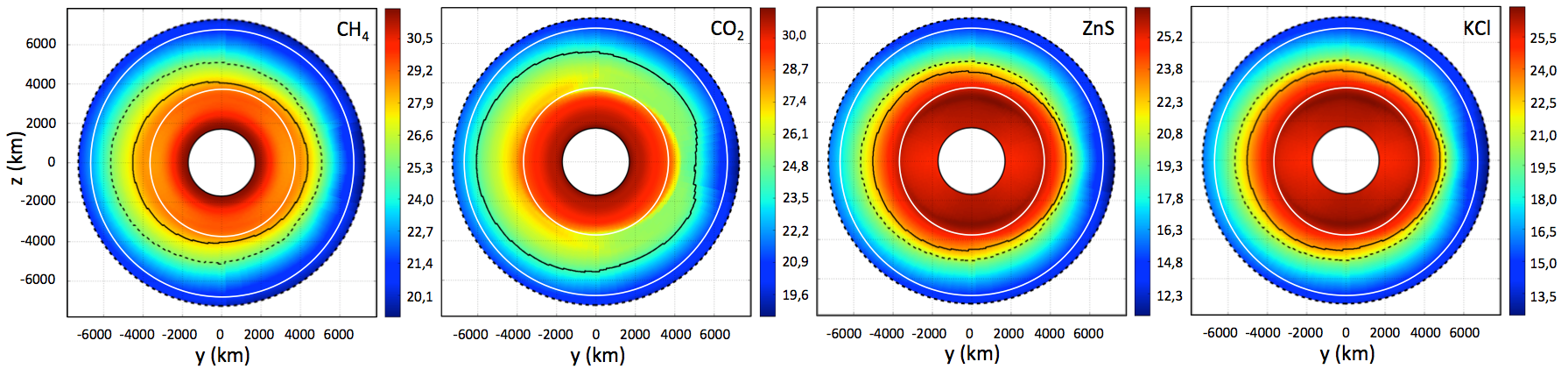}
      \caption{Decimal logarithm of the column density of some molecules in the star-observer direction for a simulation of the atmosphere of GJ\,1214\,b by \citet{CMM15}. See the caption of \fig{fig.abs} for details and notations.}
         \label{fig.abs2}
    \vspace{0.5cm}

\end{figure*}

\begin{figure*}[h]
  %  \includegraphics[width=0.95\linewidth]{transmittance_maps.png}
  %\sidecaption\includegraphics[width=12.5cm]{Trans.png}
  \includegraphics[width=0.95\linewidth]{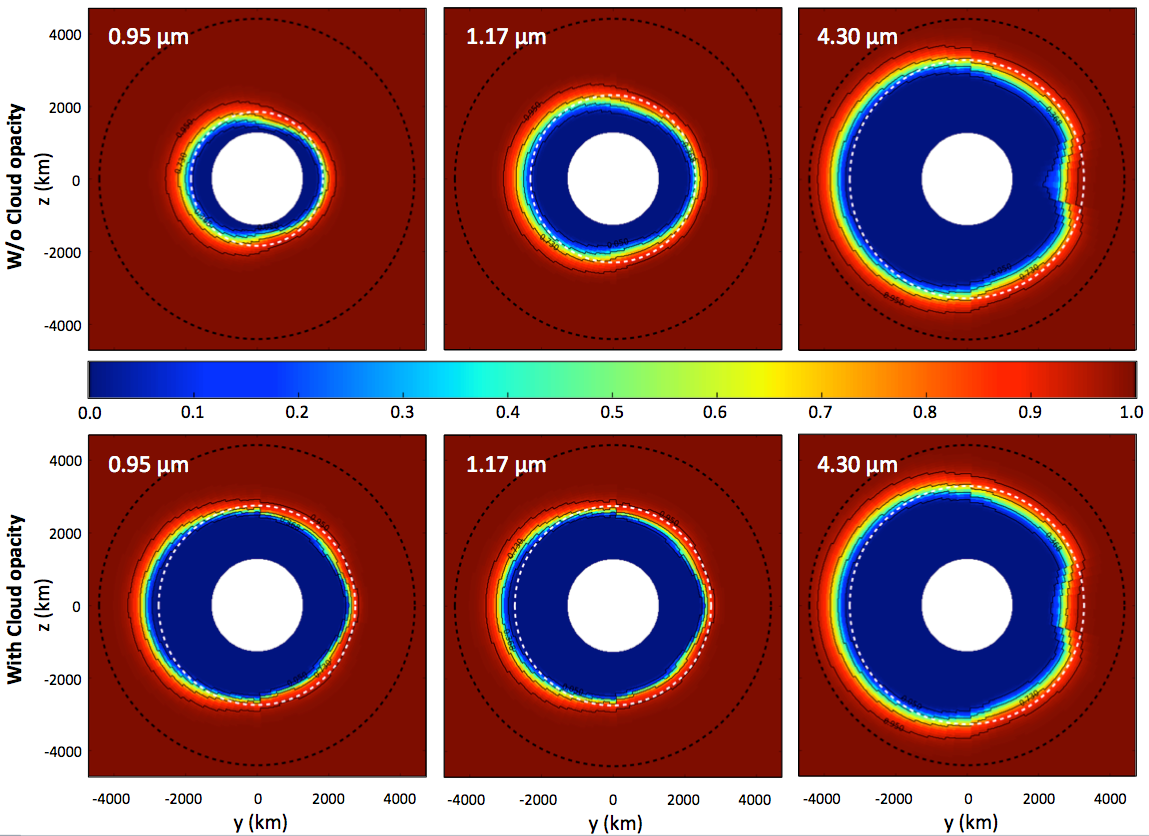}
      \caption{Transmittance maps obtained with the simulation of the atmosphere of GJ\,1214\,b depicted in \fig{fig.abs}. The region shown between the inner white disk and the outer black dotted circle is an enlargement of the region delimited by white circles in \fig{fig.abs}. The white dotted circle corresponds to the effective radius: the radius of the opaque disk resulting in the same overall absorption as the simulation over a homogeneous stellar disk. The maps in the upper row are obtained with Mie scattering turned off in order to show the effect of clouds. Clouds dominate near $0.95~\mu$m, methane is the main gaseous absorber at $1.17~\mu$m, and carbon dioxide at $4.30~\mu$m.}
         \label{maps1214}
   \end{figure*}
%%%%%%%%%%%%%%%%%%%%%%%%%%%%%%%%%%%%%%%%
%%%%%%%%%%%%%%%%%%%%%%%%%%%%%%%%%%%%%%%%

\section{The case of GJ~1214b as an illustration of Pythmosph3R capabilities}\label{Application}

In this section, we apply Pythmosph3R to GCM simulations of the atmosphere of GJ\,1214\,b to illustrate the possibilities of the code. This also allows us to show examples of horizontal inhomogeneities and some of their effects on transmission spectra. Despite the flat spectra currently obtained with WFC3/HST, possibly due to high-altitude clouds/aerosols \citep{KBD14}, GJ\,1214\,b is one of the rare known targets that is not a gas giant but still offers a favourable configuration for transmission spectroscopy  : the vicinity to the Sun (13~pc), a low-density implying the presence of an atmosphere, a short period (1.58\,d), a red dwarf host and an expected high scale height to radius ratio (see \fig{Cells}). Modelling the formation, distribution and spectral signature of aerosols motivated the use of a 3D atmospheric model of the atmosphere \citep{CMM15} that we used to compute transmission maps and spectra with Pythmosph3R.

%Indeed, as will be shown hereafter, the horizontal inhomogeneities in the model atmosphere are sufficient to create features in the modeled spectrum that are greater than the anticipated photon noise for the planet over one transit with JWST. Surprisingly, these differences do not only come from the horizontal variations of temperature and composition along the limb (e.g. from the leading to the trailing limb), but also from inhomogeneities along the path of each ray as it enters through the dayside part of the limb and leaves through the nightside. 

%Therefore, we first briefly describe some features of the 3D simulation used, then we discuss and quantify the effects of the various types of horizontal inhomogeneities discussed above. This quantification is performed by comparing the spectrum generated by our 3D tool with spectra generated by simpler 1D or (1+1)D models that will be described below. Finally, we will discuss a way to identify and measure these inhomogeneities.

\subsection{Input 3D atmospheric model}

We use the 100\,$\times$ solar metallicity simulation of \citet{CMM15} made with the LMDZ generic global climate model. The simulation has a 64 $\times$ 48 horizontal resolution with 50 layers equally spaced in log pressure, spanning 80 bars to $\sim$ 0.5\,Pa.  An important assumption in this simulation is the local chemical equilibrium : in each cell, the composition is imposed by the local temperature, pressure and elemental composition. As the simulation assumes a circular orbit with null obliquity and a synchronized rotation, the state of the atmosphere after convergence is fairly stable, exhibiting only stochastic variations. We use an arbitrary timestep to produce synthetic spectra. The thermal structure is shown in \fig{Tmap_GJ} and the absorbers/clouds distribution in \fig{fig.abs}.  We redirect the reader to \citet{CMM15} for more details on the model.

\subsection{Transmittance maps}
In order to understand the global transmission spectrum Pythmosph3R offers the possibility to draw transmittance map in any spectral bin of the spectrum. Viewing the azimuthal and vertical inhomogeneities provides some important piece of information to interpret spectral features.  

\fig{maps1214} shows transmittance maps for GJ\,1214\,b simulated atmosphere, with and without the effect of clouds. Temperature is directly or indirectly at the origin of the inhomogeneities. Colder regions at the western terminator are less extended vertically and locally induce a smaller absorption radius. As the simulation imposes a local chemical equilibrium, colder regions are also poorer in CO$_2$ and richer in CH$_4$, affecting the transmittance distribution in the displayed CO$_2$ ($14.9$ and $4.3~\mu$m) and CH$_4$ ($1.17~\mu$m) bands. In the $14.9~\mu$m CO$_2$ bands another thermal effect comes from the strong temperature dependency of the absorption cross-section, which strongly enhances the transmittance inhomogeneities compared with the $4.3~\mu$m band, much less sensitive to the temperature. The effect of clouds is the strongest at $0.95~\mu$m although they mask some chemical inhomogeneities at high pressure.

\subsection{Comparison with averaging methods based on 1D models}

%%%%%%%%%%%%%%%%%%%%%%%%%%%%%%%%%%%%%%%%
%%%%%%%%%%%%%%%%%%%%%%%%%%%%%%%%%%%%%%%%
\begin{figure*}[p]
   \centering
   \includegraphics[width=0.85\linewidth]{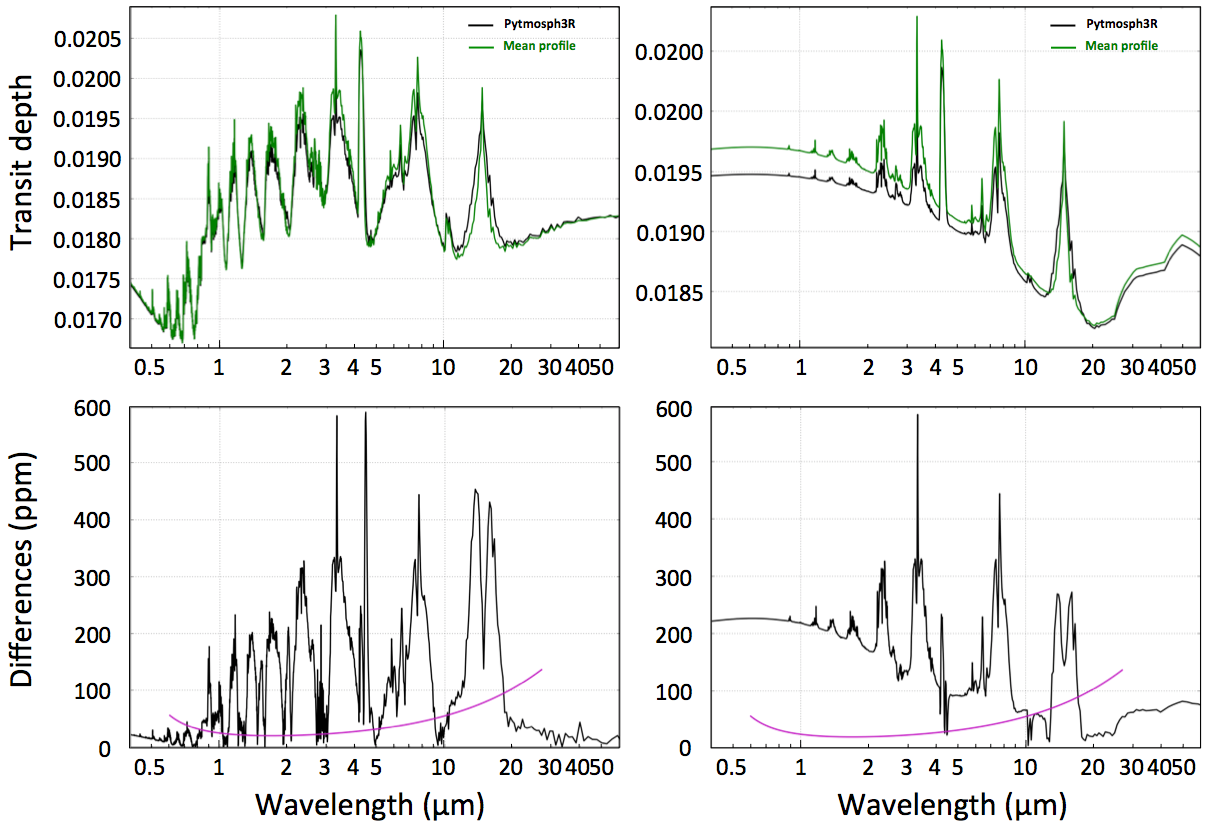}
      \caption{Transit spectra computed from a 3D simulation of GJ\,1214\,b atmosphere by \citet{CMM15} with Pythmosph3R (black) and a \textit{mean profile} approximation (see text).  Spectra are shown with (right) and without (left) the radiative effects from the $KCl$ and $ZnS$ clouds. In both cases the atmosphere is extrapolated vertically beyond the top of the simulation. Cloud particle radius is fixed to $0.5$ $\mu m$. The plots on the lower part show the difference between the two methods. The purple line indicates the photon noise with a JWST aperture, an exposure time of twice the transit duration and a (low) resolution of R = 100.}
         \label{Spec3}
 %%%%%%%%%%%%%%%%%%%%%%%%%%%%%%%%%%%%%%%%
%%%%%%%%%%%%%%%%%%%%%%%%%%%%%%%%%%%%%%%%

%%%%%%%%%%%%%%%%%%%%%%%%%%%%%%%%%%%%%%%%
%%%%%%%%%%%%%%%%%%%%%%%%%%%%%%%%%%%%%%%%
\vspace{0.5cm}
   \includegraphics[width=0.85\linewidth]{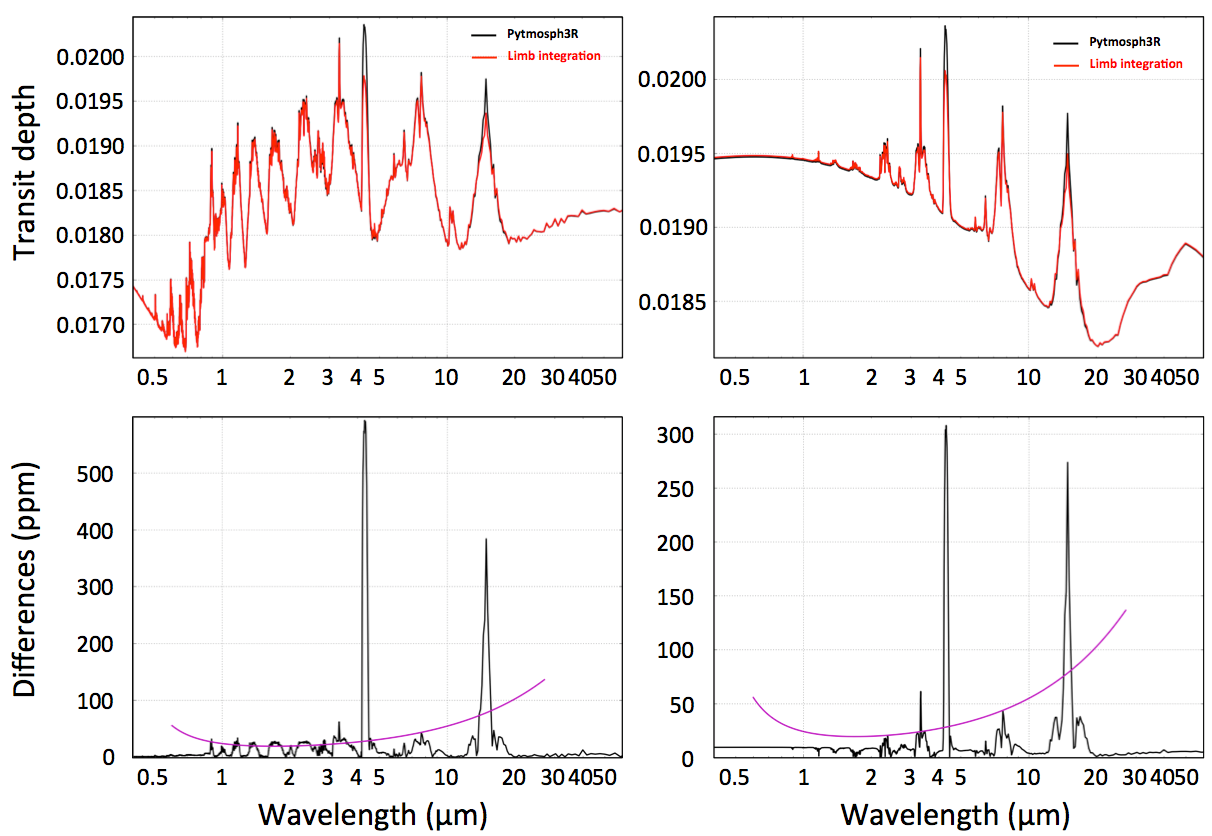}
      \caption{Same as \fig{Spec3} but the Pythmosph3R spectra (black) are compared here with those obtained with the \textit{limb integration} approximation (red).}
         \label{Spec}
   \end{figure*}
%%%%%%%%%%%%%%%%%%%%%%%%%%%%%%%%%%%%%%%%
%%%%%%%%%%%%%%%%%%%%%%%%%%%%%%%%%%%%%%%%

In the absence of a code like Pythmosph3R, producing a transmission spectrum from a 3D simulation using a radiative transfer model based on a horizontally-homogeneous atmospheric profile implies an average of some kind. 

\subsubsection{Mean profile}

One method consists in averaging first the atmospheric quantities and then computing a transmission spectrum. One single atmospheric profile (temperature, pressure, chemical abundances and cloud properties as a function of altitude) is obtained by averaging the $2N-2$ profiles found on the terminator, weighted by the fraction of azimuth covered by each cell ($N$ being the number of latitude points on the simulation grid). The transmission spectrum is then computed assuming that this atmospheric profile covers the whole planet. \fig{Spec3} shows the comparison between spectra resulting from this \textit{mean profile} method and those generated by Pytmosph3R.

The difference is in large part due to the lower temperature patch at the morning terminator near the equator (west of the substellar point) which is due to the equatorial jet bringing cold air from the nightside there (see \fig{Tmap_GJ}). This difference is considerably larger than the expected photon noise with JWST over most of the spectrum making this method clearly inadequate. This, in a sense, already highlights why 1D retrieval methods may be biased when interpreting the spectra of real atmospheres. 

\subsubsection{Limb integration}

A better technique that is usually applied, for instance by \citet{CMM15}, \citet{WAA17}, \citet{PLB18}, or \citet{LMM18},  involves computing in a first step $2N-2$ transmission spectra assuming a horizontally-homogeneous atmosphere, one for each atmospheric column found on the terminator. Then the total transmission spectrum is calculated as an average of these intermediate spectra, weighted by the fraction of azimuth covered by each column.
\fig{Spec} compares spectra obtained with this \textit{limb integration} technique and shows the comparison between spectra resulting from this approach (red line) and from Pytmosph3R (red line).

A rapid comparison of \figs{Spec3}{Spec} clearly shows that \textit{limb integration} performs much better than the mean profile approach. However, there still remains significant discrepancies throughout the spectrum and especially in some molecular bands. By definition, these differences only come from the atmospheric inhomogeneities \textit{along} the path of the ray. They are due to the effect of the
\begin{itemize}
    \item[$\bullet$] Day to night temperature gradient: as the dayside is hotter than the night side, the vertical extent of the atmosphere changes along the ray. As shown in \sect{daynightretrieval}, this causes a net increase in absorption visible in the water bands (see lower panels of \fig{Spec}). Although this effect is on the order of the photon noise for a single transit for the (relatively) cold atmosphere of G\,J1214\,b, it can strongly affect the retrieval of the properties of hotter planets as demonstrated in the next section. 
    \item[$\bullet$] Day to night compositional gradient: for the absorption bands due to absorbers with an heterogeneous distribution (like CO$_2$ is our case, which absorbs prominently at 4.5 and 15\,$\mu$m) a change of composition along the line of sight creates a signal that is much greater than the expected noise and that cannot be modelled by the \textit{limb integration} method. See below for the possible causes of such a compositional gradient. Although it is the most prominent effect in our GJ\,1214\,b  model, the parameter space to cover to fully quantify it is large and will have to wait a future study.  
    \item[$\bullet$] Day to night asymmetry of the cloud distribution: the temperature change at the terminator, in turn, allows us to expect changes in the properties of the clouds there \citep{LDH16}. This will also be looked at in a future study.
\end{itemize}

\subsection{Comments on the possible causes of compositional heterogeneities}

In the model of GJ\,1214\,b we use, the concentrations of CO, CO$_2$ and CH$_4$ are computed assuming chemical equilibrium. For such atmospheres and cooler ones, the 3D variations of these species are thus strongly overestimated. While the hottest regions of the atmosphere (above $\sim 1000$~K) are expected to reach equilibrium faster than typical dynamic timescales, it is not the case in the coldest layers (below $\sim 700$~K) probed by transmission, where endothermic reaction become extremely slow. As a consequence, thermochemistry is expected to produce a more homogeneous composition controlled by the hottest/deepest regions.

However, more irradiated planets must have overall hotter atmospheres spanning a larger range of temperatures (due to shorter radiative timescales) with very different equilibrium compositions and  kinetics fast enough to main local equilibrium \citep{agundez2014}. These atmospheres are expected to exhibit the strongest horizontal variations of temperature and composition. \citet{PLB18} recently showed that water itself may be dissociated on the dayside of some hot-Jupiters and recombine close to their terminator. In addition, UV-driven photochemistry may create additional heterogeneities by allowing some reactions to take place on the dayside of the planet only.

We can thus expect hot planets to exhibit strong day to night compositional gradients that may become a dominant issue in retrieving atmospheric properties through transmission spectroscopy. Quantifying the stellar irradiation at which these effects become significant will need further modelling.

\begin{figure*}[tbp] %  figure placement: here, top, bottom, or page
 \centering
\subfigure{ \includegraphics[scale=.55,trim = 0cm .cm 3.cm 0.cm, clip]{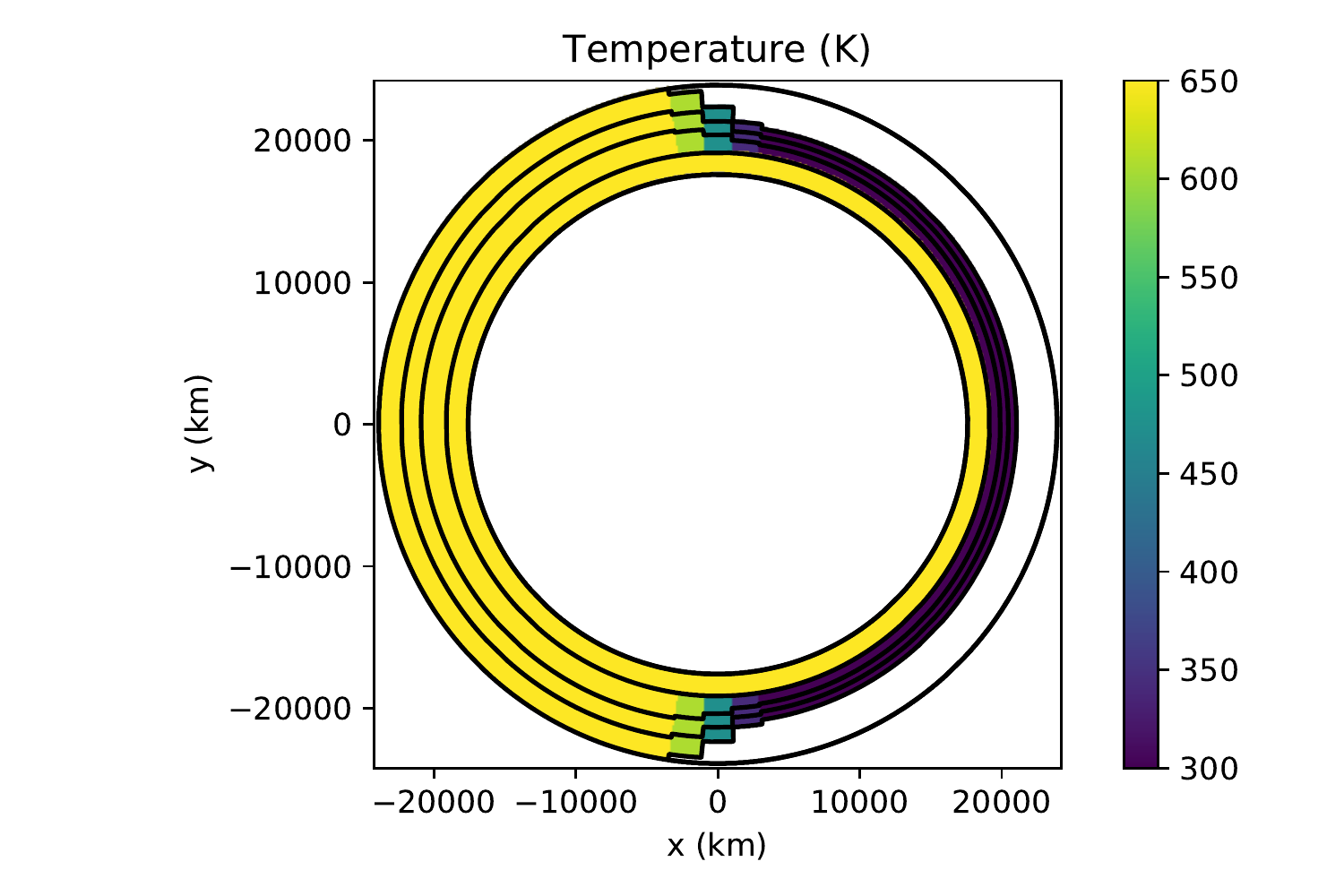} }
\subfigure{ \includegraphics[scale=.55,trim = 4.cm .cm 3.cm 0.cm, clip]{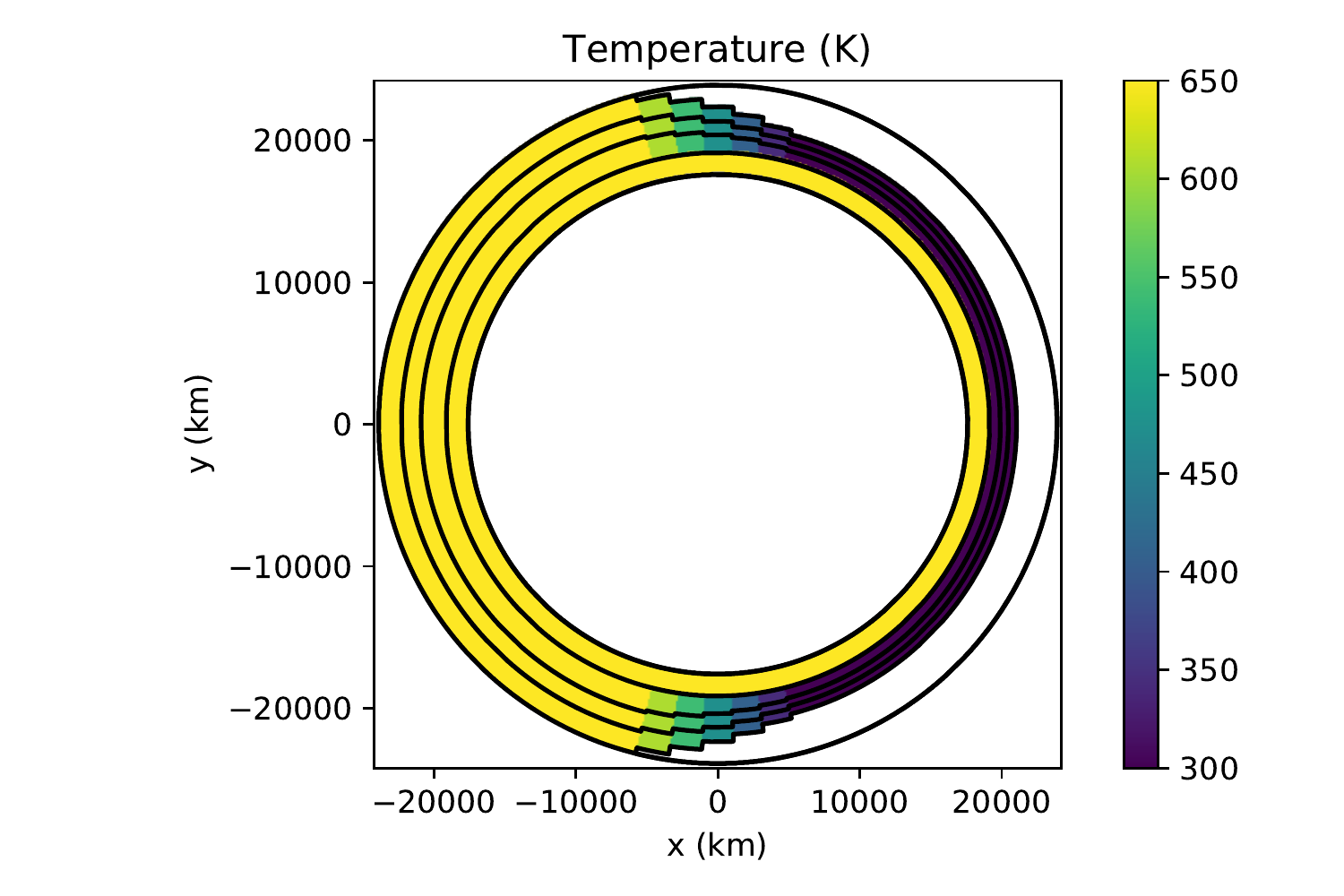} }
\subfigure{ \includegraphics[scale=.55,trim = 4.cm .cm .cm 0.cm, clip]{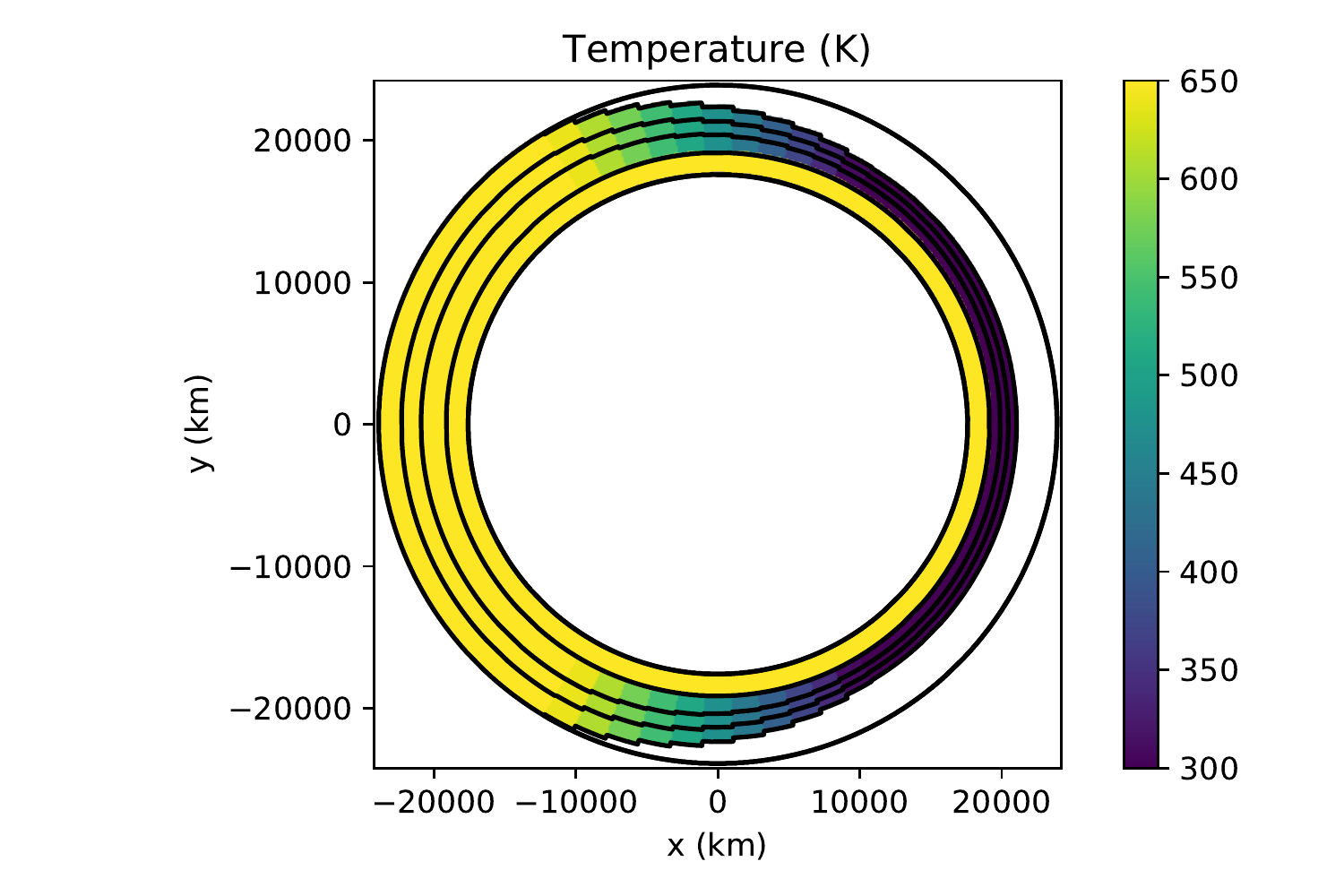} }
%\subfigure{ \includegraphics[scale=.55,trim = 3.cm .cm .cm 0.cm, clip]{figDuoT/TempEquatorMap_GJ1214brad_levs.pdf} }
\caption{
These color maps show the temperature distribution of our model atmospheres in a plane perpendicular to the terminator for three different transition angles between a 650\,K dayside and a 300\,K nightside (from left to right, the day-night transition angle is $\openAngle$=15$^\circ$, 30$^\circ$, and 60$^\circ$). The inner white circle represents the inner part of the planet with a radius assumed to be equal to the one of GJ\,1214\,b (17.600\,km). The star is on the left and the observer on the right on the $y=0$ line. From center outward, the 5 solid lines are respectively the 10$^6$, 10$^3$, 1, 10$^{-2}$, and 10$^{-4}$\,Pa pressure levels. Below the 10$^3$\,Pa level, the atmosphere is assumed to efficiently redistribute heat and is horizontally isothermal. These maps are \textit{to scale} and show that the dayside is noticeably more extended than the nightside.  }
 \label{TEquMap_GJ}
\end{figure*}

\section{Effect of Day/night side temperature differences on retrieval}\label{daynightretrieval}

As visible in \fig{Cells}, the region probed in primary transit is much larger than is usually acknowledged, especially on hot and/or low gravity objects. So during transit, we are not only probing a thin plane -- the so-called terminator -- but an area that can extend significantly on both the day and night sides.
Because these two parts of the planets are expected to exhibit quite different temperatures (as visible in \fig{Tmap_GJ}), it seems important to quantify the extent of the imprint of this temperature inhomogeneity on the transit spectrum of the planet, and how it will affect any attempt to retrieve the temperature at the terminator.

To answer these questions, we conduct a simple experiment. For two prototypical planets (respectively based on GJ\,1214\,b and HD\,209458\,b), we build idealized 3D atmospheric structures that are symmetric about the star-planet axis but that continuously go from a high temperature $\Tday$ on the dayside to a lower one, $\Tnight$, on the nightside. The atmosphere is assumed to have a \textit{uniform composition} to enable us to concentrate on thermal effects. We then simulate the transit spectrum and try to invert it. Finally, the retrieved temperature is compared to the input one.

\subsection{Parametrization of the atmospheric structure}

 The structure that we chose for the atmosphere is inspired by the 3D simulations of \citet{CMM15} for GJ\,1214\,b whose temperature distribution is shown in \fig{Tmap_GJ}. Its most salient feature is the continuous transition in temperature between the day and night side. For sake of simplicity, we assume that this transition occurs linearly over a region that is parametrized by its opening angle --- hereafter called the transition angle ($\openAngle$).
 
 Moreover, as is well known and further exemplified by \fig{Tmap_GJ}, the atmosphere of gaseous exoplanets is usually well mixed at depth. The temperature is thus assumed to be uniform below a pressure level $\piso$ (that will taken to be 10\,mb as in the simulation, although some models do predict inhomogeneities to persist at deeper levels). 

In summary, for a given location identified by its pressure, $\press$, and the (possibly negative) local solar elevation angle, $\thstar$, the temperature is given by
\balign{
\left\{\begin{array}{cc}\press > \piso & T=\Tday
\\
\press < \piso & \left\{\begin{array}{cl}2\thstar > \openAngle & T=\Tday \\
\openAngle > 2\thstar > -\openAngle & T=\Tnight+ \left(\Tday -\Tnight\right) \frac{\thstar+\openAngle/2}{\openAngle} \\
- \openAngle > 2\thstar & T=\Tnight\end{array}\right.\end{array}\right.
}

Examples of such idealized atmospheric structures are shown in \fig{TEquMap_GJ} for three different day-night transition angles. We chose to show the structures that are most representative of the real GJ\, 1214\,b case to allow for a direct comparison with \fig{Tmap_GJ}. For brevity we will also refer hereafter to the "uniform case": this stands for a case where the whole upper atmosphere has a uniform temperature equal to the one at the terminator that serves as a comparison. Therefore, in this case,
\balign{
\left\{\begin{array}{cl}\press > \piso & T=\Tday
\\
\press < \piso & T=(\Tday+\Tnight)/2\end{array}\right. .
}

Once our four parameters -- $\Tday$, $\Tnight$, $\openAngle$, and $\piso$ -- have been chosen, the 3D atmospheric structure is integrated from the surface of the planet that is assumed to be the 10\,bar level. Hereafter, we call the radius of this isobar $\Rp$, and assume that it contains most of the mass of the object so that the gravity above this level only depends on the altitude.

The reason we are not performing such test on idealized structures and not more realistic ones fro a GCM is that we want to isolate the effect of the day-night heterogeneities. In a GCM simulation, there would also be vertical and equator-pole temperature variations that would preclude the identification of a single effect.

\begin{figure*}[tbp] %  figure placement: here, top, bottom, or page
 \centering
\subfigure{ \includegraphics[scale=.7,trim = 0cm .cm 1.8cm 0.cm, clip]{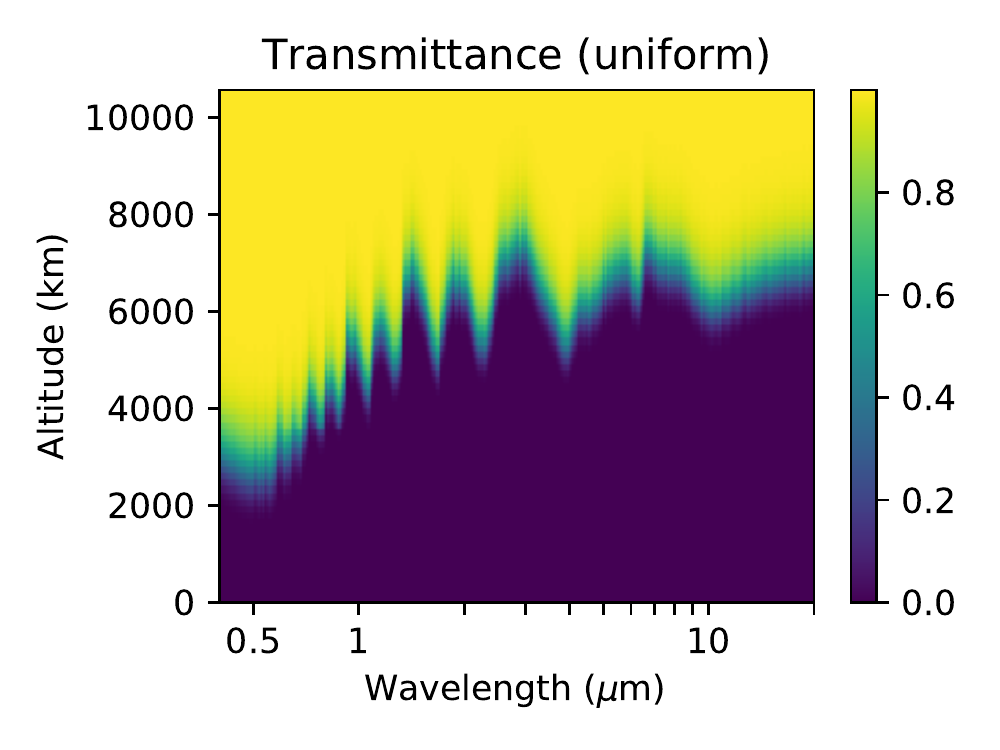} }
\subfigure{ \includegraphics[scale=.7,trim = 2.cm .cm 0.2cm 0.cm, clip]{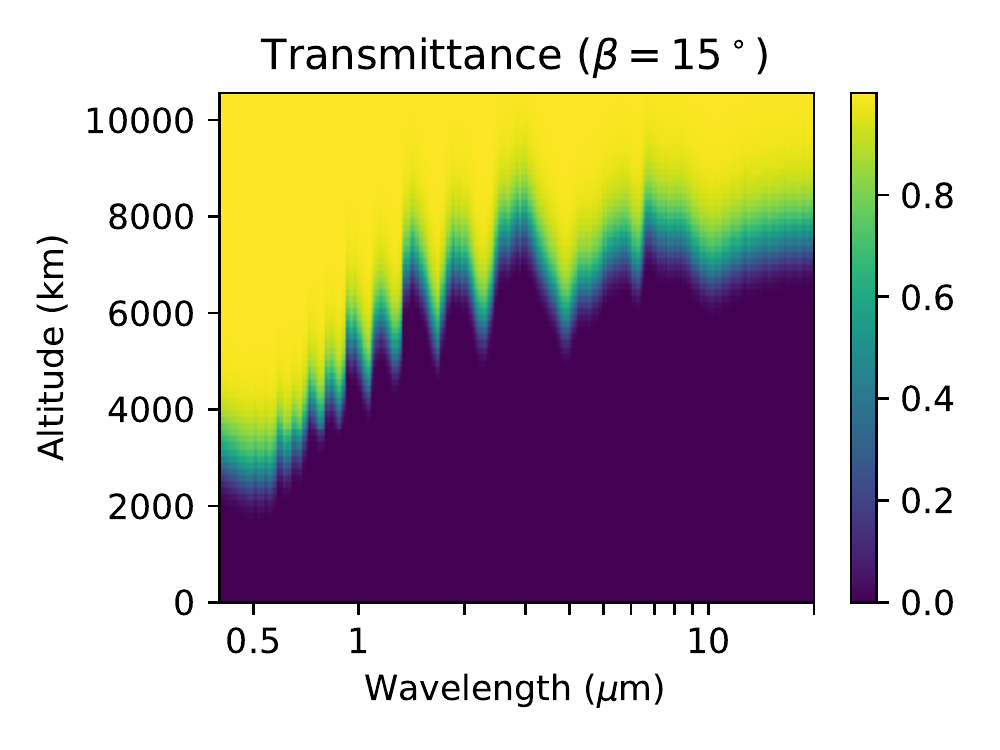} }
\subfigure{ \includegraphics[scale=.7,trim = 2.cm .cm .cm 0.cm, clip]{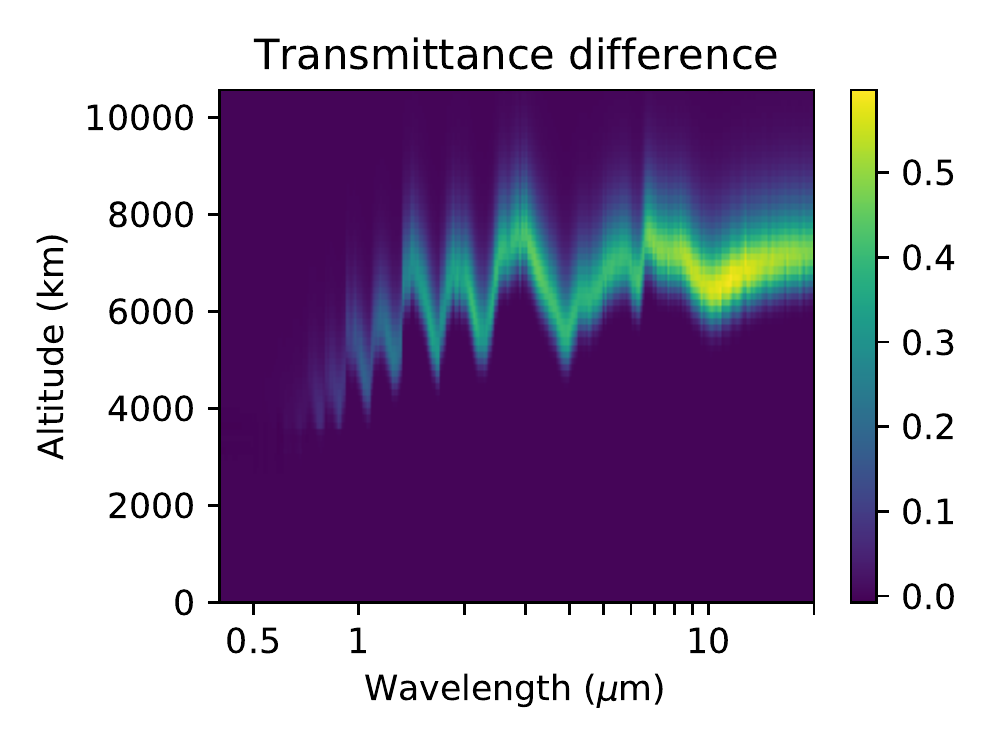} }
\caption{
Maps of the spectral transmittance of clear atmosphere models of HD\,209458\,b as a function of wavelenth and altitude. The parameters are $\Tday=1800$\,K, $\Tnight=1000$\,K, and $\piso=10\,$mb, that are representative of the real planet \citep{PSL13}. A low transmittance -- blue shades -- is representative of opaque regions deep down and a transmittance near unity -- yellow -- is representative of the transparent upper atmosphere. The left panel shows the case with a uniform upper temperature of 1400\,K, and middle one, the case with a $\openAngle=15^\circ$ transition region between the day and night side. The higher opacity of the middle case is further highlighted by the right panel that shows the map of the transmittance difference between the left and the middle case. The altitude of the top of the isothermal region is $\sim 3\,800\,$km.
 }
 \label{TransMAP}
\end{figure*}

\begin{figure*}
   \centering
 \subfigure{ \includegraphics[scale=.23,trim = 0cm .cm 0cm 0.cm, clip]{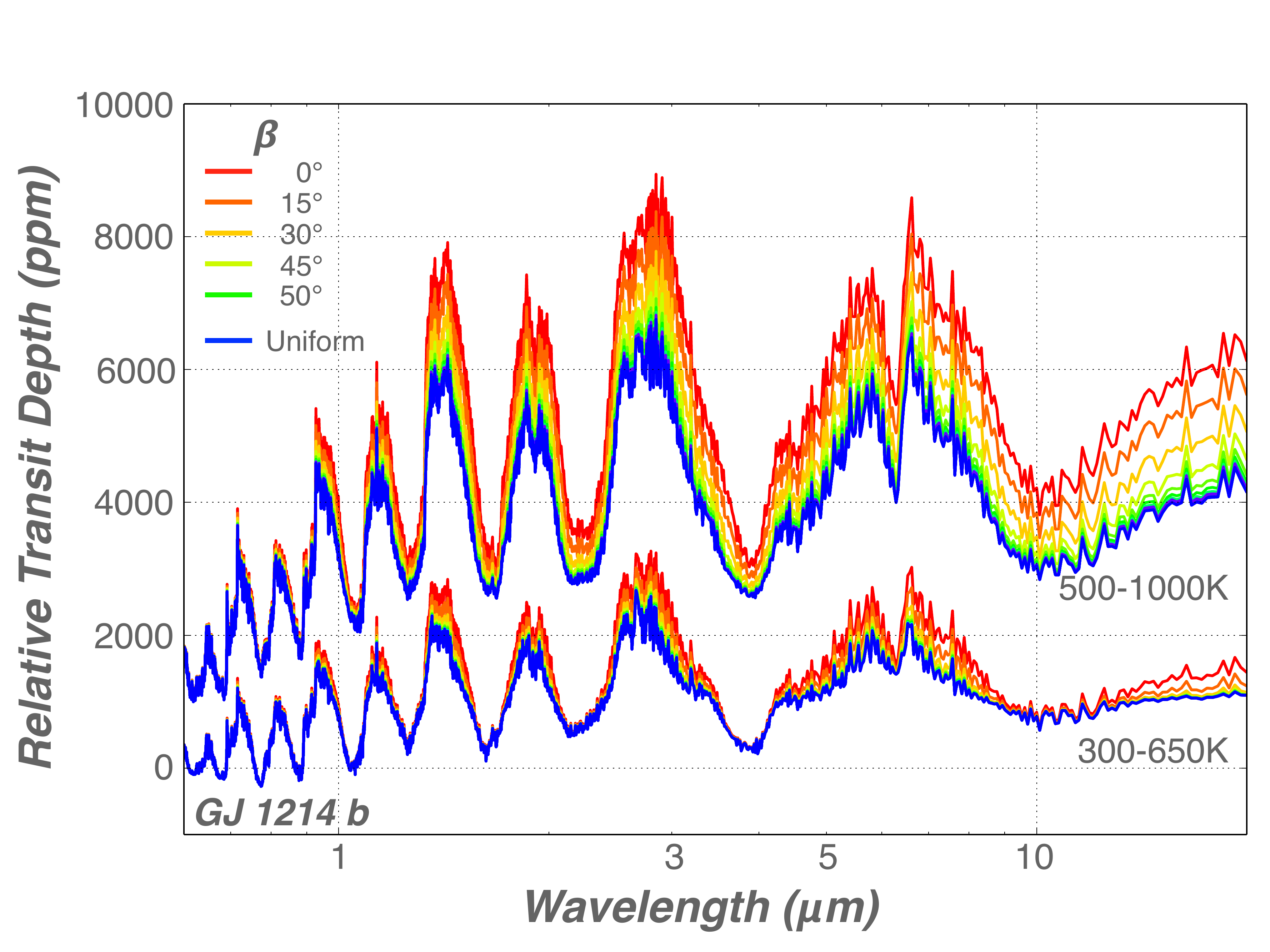} }
 \subfigure{ \includegraphics[scale=.23,trim = 1.8cm .cm 0cm 0.cm, clip]{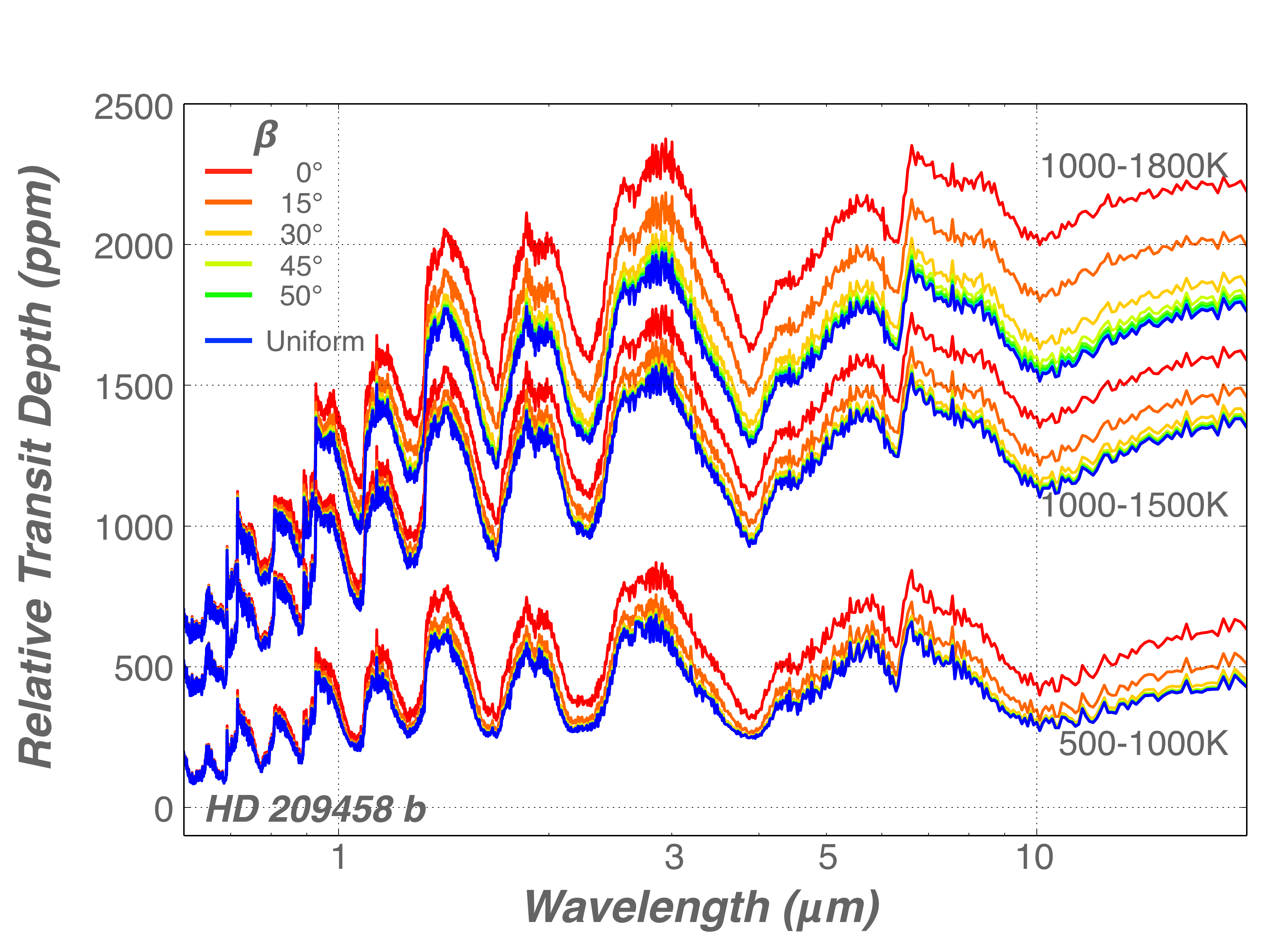} }
      \caption{Spatially integrated spectra of the relative transit depth (in ppm) expected for GJ\,1214\,b (left) and HD\,209458\,b (right) as a function of the day-night transition angle ($\openAngle$). The color goes from red to green when $\openAngle$ goes from 0$^\circ$ to 50$^\circ$ every 15$^\circ$.  The blue curve is the reference uniform case. Spectra with larger $\openAngle$ are undistinguishable from the uniform case. The different labeled groups of curves are for our various sets of day-night temperatures. For each $(\Tday-\Tnight)$ group, the temperature at the terminator of all the models -- including the uniform one -- is $(\Tday+\Tnight)/2$. It can be seen that the transit depth increases monotonically when $\openAngle$ decreases. }
         \label{spectrumdiff}
   \end{figure*}

\subsection{Effect of the day/night temperature difference on the transmission spectrum}

One might naively except that because of the symmetry of our temperature distribution, the contributions of the hot and cold sides should cancel out. This is however not the case, as can be seen in \fig{TransMAP}.

Indeed, by comparing the transmittance map for a given planet in a uniform case, or with a day-night temperature gradient, we directly see that the opaque region extends higher up in the latter case. The greater scale height on the dayside is not compensated by the lower one on the night side. 

This can be understood easily using a slightly modified version of the analytical model of \citet{Guigui} or \citet{VCM14} where we separate the atmosphere into two hemispheres that differ only through their temperatures -- and thus atmospheric scale heights ($\Hday$ and $\Hnight$) -- above the pressure level $\piso$ which is located at an altitude $\ziso$ above the reference radius of the planet $\Rp$. Below $\ziso$, the scale height is the same everywhere ($\Hdeep$). In essence, this corresponds to the case described above in the limit where $\openAngle \rightarrow 0.$

 We follow the notations in \fig{schematic_transit} and the formalism in \app{app:guillotmodel}. The difference here is that the scale height variations entail that the number density at a given altitude is 
\balign{
\ngas(\z)=\left\{\begin{array}{ll}
\ngasref e^{-\frac{\z}{\Hdeep}} & \z < \ziso 
\\
 \ngasref e^{-\frac{\ziso}{\Hdeep}}e^{-\frac{\z-\ziso}{H_\mathrm{i}}}
 &\z > \ziso 
\end{array}\right. ,
}
where i is either day or night depending on the hemisphere. 

In the limit where all the altitudes in the atmosphere are small compared to $\Rp$, the slant optical depth is the sum of the day and night side contribution, yielding
\balign{
\left. \tautr \right|_{\zt>\ziso}&=\overbrace{\int_{-\infty}^0 \ngas \sigmol \d \x }^{\mathrm{day}}  +\overbrace{\int_0^{\infty} \ngas \sigmol \d \x}^{\mathrm{night}}
\nonumber\\
&= \sigmol \ngasref e^{-\frac{\ziso}{\Hdeep}} \left(\sum_{\mathrm{i=day}}^{\mathrm{night}}e^{-\frac{\zt-\ziso}{H_\mathrm{i}}}\int_0^\infty e^{-\frac{\x^2}{2 \Rp H_\mathrm{i}}}\d \x \right)  \nonumber\\
&=\sigmol \ngasref e^{-\frac{\ziso}{\Hdeep}}  \left(  e^{-\frac{\zt-\ziso}{\Hday}}\sqrt{\frac{\pi \Rp \Hday}{2}}  + e^{-\frac{\zt-\ziso}{\Hnight}}\sqrt{\frac{\pi \Rp \Hnight}{2}} \right),
}
if $\zt>\ziso$, and 
\balign{
\left. \tautr \right|_{\zt<\ziso}
&=\sigmol \ngasref \left( \int_{-\sqrt{2(\ziso-\zt)\Rp}}^{\sqrt{2(\ziso-\zt)\Rp}} e^{-\frac{\zt}{\Hdeep}}e^{-\frac{\x^2}{2 \Rp \Hdeep}}\d \x \right. \nonumber\\
& \hspace{0.cm}\left. +e^{-\frac{\ziso}{\Hdeep}}\sum_{\mathrm{i=day}}^{\mathrm{night}}\int_{\sqrt{2(\ziso-\zt)\Rp}}^\infty e^{-\frac{\zt-\ziso}{\H_\mathrm{i}}}e^{-\frac{\x^2}{2 \Rp \H_\mathrm{i}}}\d \x\right)
\nonumber\\
&=\sigmol \ngasref \left( e^{-\frac{\zt}{\Hdeep}}\sqrt{2 \pi \Rp\Hdeep}\,\mathrm{erf}[\sqrt{\frac{\ziso-\zt}{\Hdeep}}] \right.\nonumber\\
&\hspace{0.cm}\left.+e^{-\frac{\ziso}{\Hdeep}}\sum_{\mathrm{i=day}}^{\mathrm{night}} e^{-\frac{\zt-\ziso}{\H_\mathrm{i}}}\sqrt{\frac{\pi\Rp\H_\mathrm{i}}{2}} \left(1-\mathrm{erf}[\sqrt{\frac{\ziso-\zt}{\H_\mathrm{i}}}]\right)\right),
}
if $\zt<\ziso$. Note that in the above equations, $\sigmol$ is the mean cross section of the gas normalized to the total density. 

Now, to see the increase in optical depth, let us divide the result above by the optical depth in the uniform case ($\tau_{\mathrm{uni}}$) given by \eq{tautransit}. This yields
\balign{
\frac{\left. \tautr \right|_{\zt>\ziso}}{\tau_{\mathrm{uni}}}=
\frac{1}{2}\left( \sqrt{1+ \dH} \, e^{\frac{\dH \dz}{1+\dH}}+\sqrt{1- \dH} \, e^{-\frac{\dH \dz}{ 1-\dH}}\right),
}
where $\dH\equiv (\Hday-\Hnight)/2$ so that $(\Hday,\Hnight)=\Hdeep(1 \pm \dH)$, and $\dz=(\zt-\ziso)/\Hdeep$. The expansion in $\dH$ gives
\balign{
\frac{\left. \tautr \right|_{\zt>\ziso}}{\tau_{\mathrm{uni}}}\approx 1+\frac{1}{8}\left(-1-4\dz+4\dz^2 \right) \dH^2 + \mathcal{O}(\dH^4).
}
First, we see that because of the symmetry of the setup, the first order term disappears. Second, it readily results that the optical depth in the heterogeneous case is larger than in the uniform case for all the rays with a tangent altitude that is $(1+\sqrt{2})/2\approx1.2$ scale heights greater than the altitude of the isothermal region. This qualitatively explains why there is little difference in the transmittance below the altitude of the isothermal region ($\sim3\,800\,$km) in \fig{TransMAP}.

Finally, when these transmittance maps are integrated vertically, we get the transmission spectra shown in \fig{spectrumdiff}. As expected, the effective altitude at which the atmosphere becomes opaque systematically increases when the horizontal thermal gradient is increased near the terminator. 

An important point is that the spectrum for the non-uniform case -- although different from the spectrum that is be obtained for a uniform atmosphere with the temperature of the terminator -- can be similar to the spectrum that would be obtained by a uniform, but globally hotter atmosphere. It is thus not surprising that a retrieval algorithm would be biased and retrieve a hotter atmosphere.

\subsection{Bias in retrieved temperatures}\label{sec:bias}

\begin{figure*}
   \centering
 \subfigure{ \includegraphics[scale=.6,trim = 0.cm .cm 0cm 0.cm, clip]{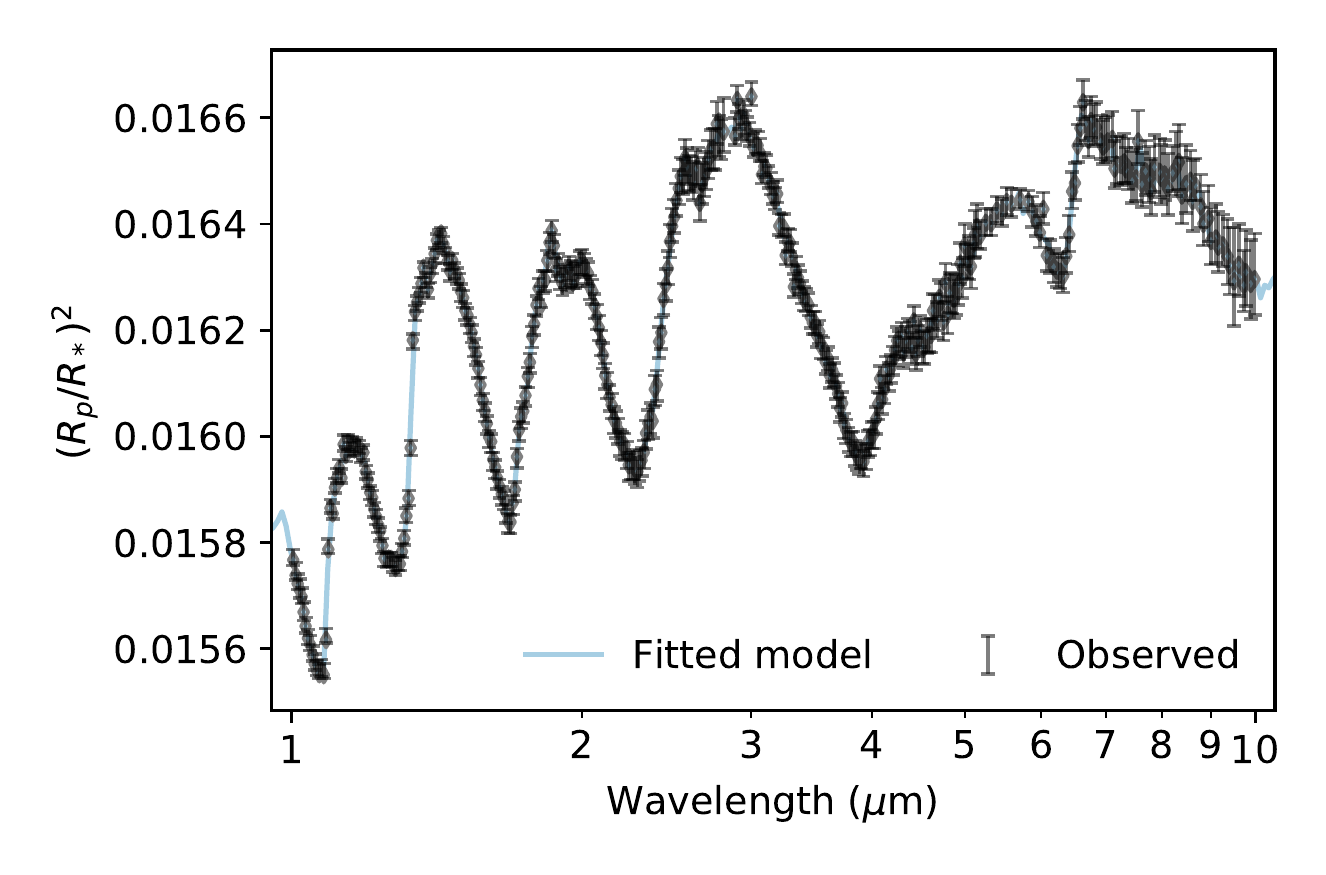} }
 \subfigure{ \includegraphics[scale=.35,trim = 0cm .cm 0cm 0.cm, clip]{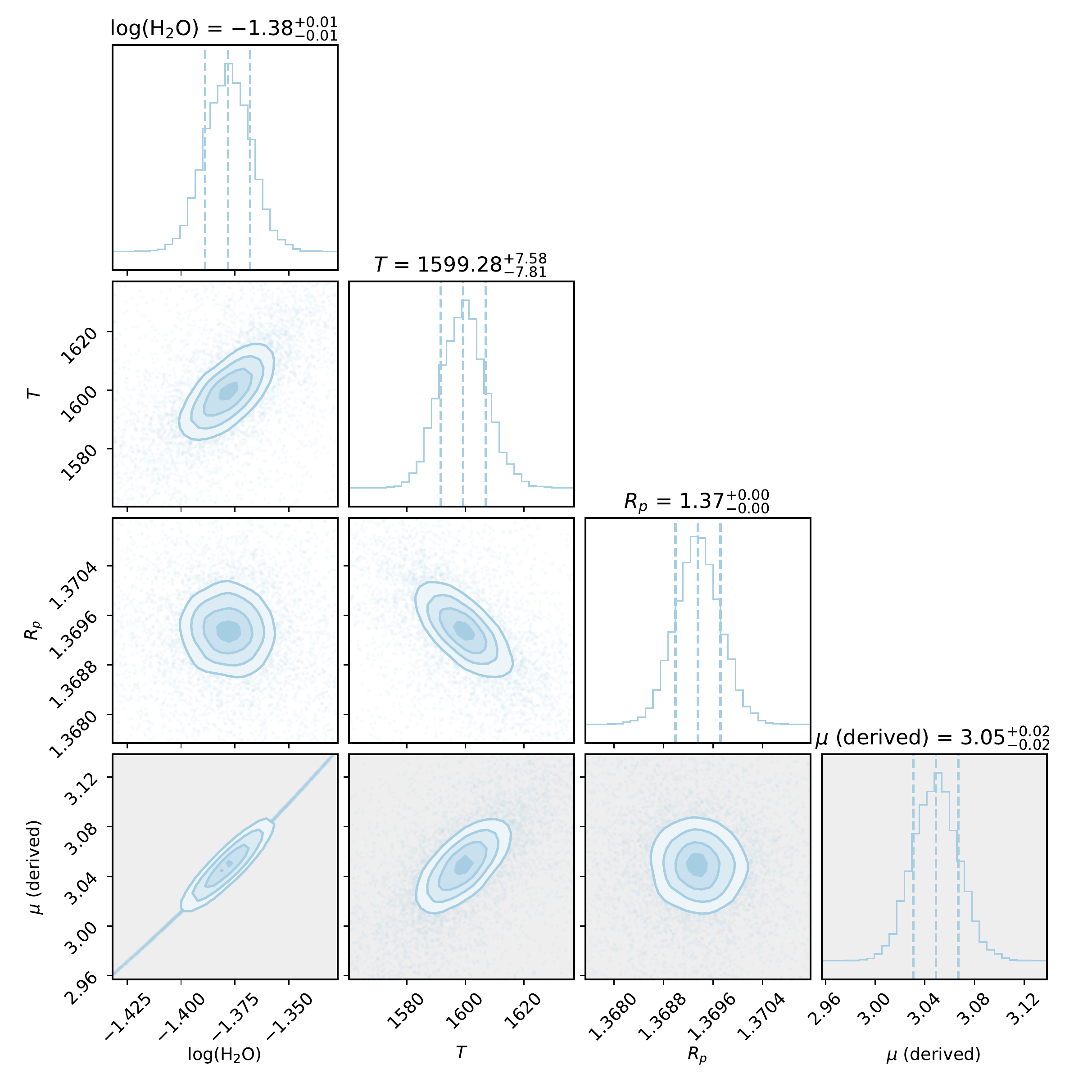} }
      \caption{ Typical result of the retrieval procedure. This specific case is HD\,209458\,b with $\openAngle=15^\circ$ and temperatures of 1800 and 1000\,K on the day and night sides respectively. The left panel is the best fit 1D spectrum (blue curve) along with the spectrum produced by our 3D tool used as input for the retrieval (black points with error bars being the 1$\sigma$ uncertainty computed with PandExo). The right panel shows the posterior distribution for the retrieved parameters. This shows that the retrieval finds an acceptable fit, which results in relatively peaked posterior distribution and small error bars on the retrieved parameters. These values are however biased as the actual terminator temperature (1400\,K) and atmospheric water abundance ($\log_{10}[\mathrm{H_2O}]=-1.3 $) are outside the range of values shown.  }
         \label{fig:posterior}
   \end{figure*}

%\begin{figure*}[tbp] %  figure placement: here, top, bottom, or page
% \sidecaption
%\includegraphics[scale=0.43]{Fig_JL/posterior_HD1000-1800_15deg.pdf}
% \caption{test}
% \label{fig:posterior}
%\end{figure*}
%

\begin{figure}[htbp] %  figure placement: here, top, bottom, or page
 \centering
\subfigure{ \includegraphics[scale=.8,trim = 0cm .8cm .cm 0.cm, clip]{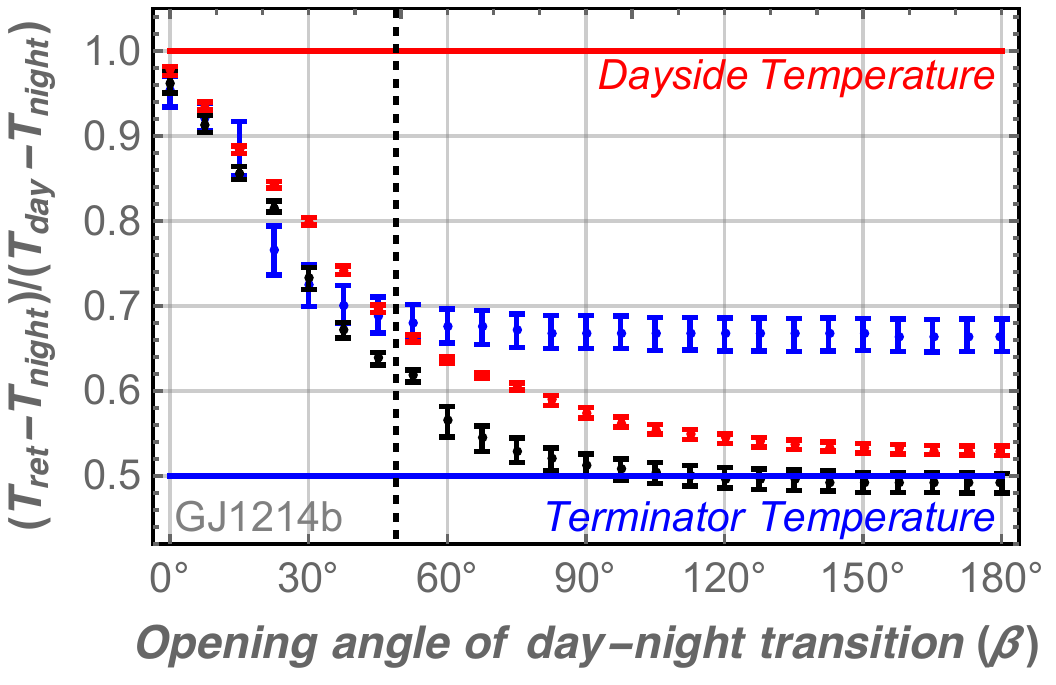} }
\subfigure{ \includegraphics[scale=.8,trim = 0cm .cm 0.cm 0.1cm, clip]{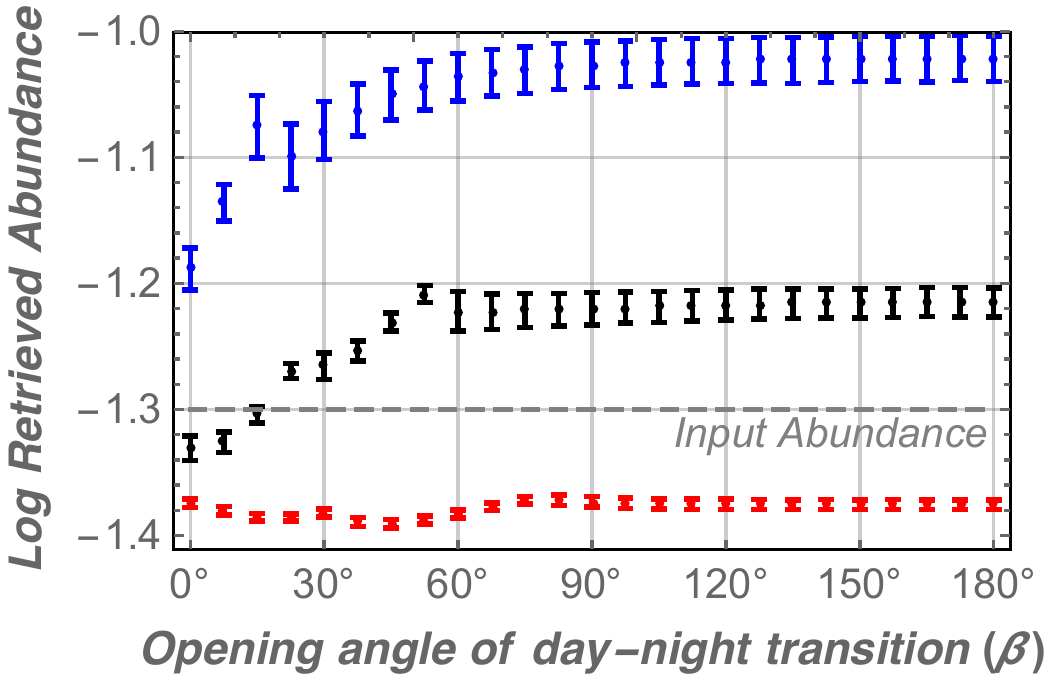} }
\caption{Retrieved temperature (top) and water abundance (bottom) as a function of the transition angle between the day and nightside of a planet with the radius and surface gravity of GJ\,1214\,b. The colours correspond to the temperature range (Blue: 300-650\,K; Black: 500-1000\,K; Red: 1000-1500\,K). To put all cases on a single diagram we show the relative retrieved temperature, $\Trel\equiv(\Tret-\Tnight)/(\Tday-\Tnight)$, so that in all cases $\Trel=0.5$ at the terminator. The retrieved temperature is systematically biased toward the dayside temperature ($\Trel\ge$0.5), especially when the transition is sharp. For comparison the estimate of the width of the limb given by \eq{geometricallimb} is shown by the vertical dotted line. }
 \label{retrievalGJ}
\end{figure}

The last step of our analysis is to actually run a 1D retrieval procedure on the transmittance spectra obtained with our parametrized 3D atmospheric structures to assess the biases entailed by such an approach. 
To do so, we use the JWST simulation tool PandExo \citep{pandexo} to simulate the expected uncertainties over a wavelength range of 1.0 - 10\,$\mu$m for the duration of a single transit. For this, we combined the simulated spectra of the NIRISS/SOSS, NIRSpec/G395M and MIRI/LRS instruments. Observed uncertainties above 10\,$\mu$m were too large to have a significant impact on retrieval results and were discarded. Finally, we binned the simulated observations to a resolution of 100 constant in wavelength. Given that we are investigating biases due to the model rather than observational noise, we here only consider the uncertainties calculated by PandExo but do not add additional noise to the mean of our simulated observations. Significant biases in the posterior distributions of retrieved parameters can occur when considering single random noise instances of the data. To correctly alleviate such noise-induced biases, one would need to combine posterior distributions of multiple noise-instance retrievals. Fortunately, \citet[sec. 5.2]{YKF18} have shown the combination of multiple noise-instance retrievals to converge to the noise-free retrieval posterior distributions, as expected from the central limit theorem. 
We have furthermore neglected the inclusion of non-Gaussian noise due to instrument or stellar systematics, as these are either not currently known and/or data set dependent. We therefore note that the retrieved parameter uncertainties presented here are theoretical lower limits.

For each model of our grid in temperature and day-night transition angle, we ran the spectrum through the TauREx retrieval software \citep{wald}.
Here we considered water as the only trace gas with absorption cross sections computed using the \citet{bt2} line list. We include Rayleigh scattering and collision-induced absorption of H$_2$-H$_2$ and H$_2$-He \citep{borysow2001,borysow2002,rothman2013} and assumed the atmosphere to be cloud-free. The vertical temperature-pressure profile was modelled to be isothermal. A typical posterior distribution for the retrieved parameters resulting from this procedure is shown in \fig{fig:posterior}, along with the input simulated spectrum and the fitted one.

\begin{figure}[htbp] %  figure placement: here, top, bottom, or page
 \centering
\subfigure{ \includegraphics[scale=.8,trim = 0cm .cm .cm 0.cm, clip]{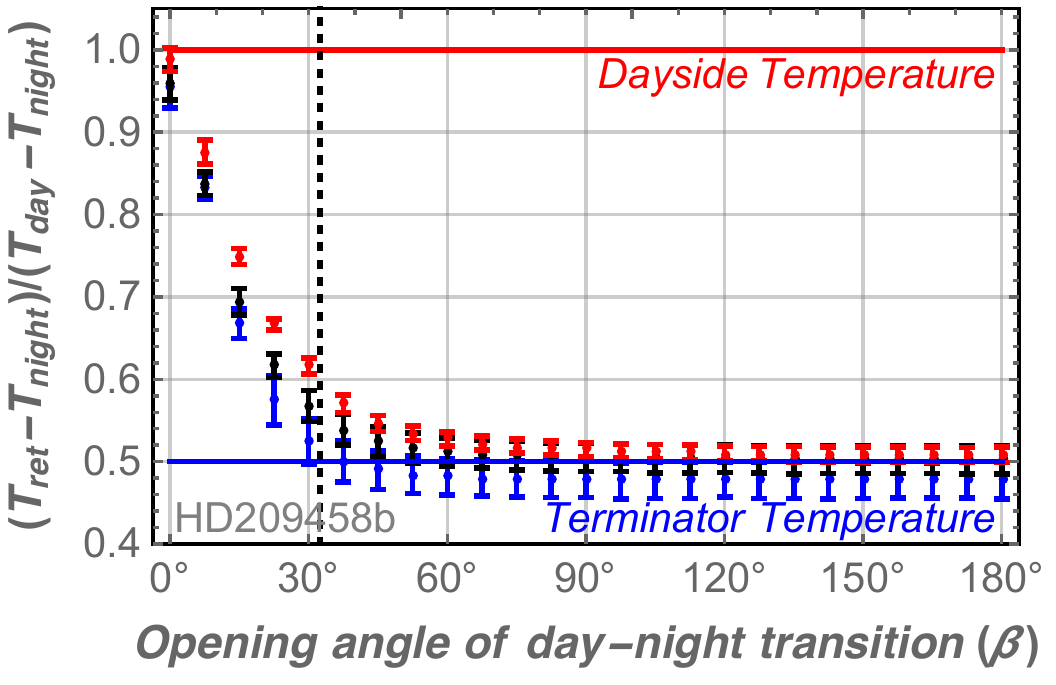} }
\caption{Dimensionless retrieved temperature as a function of the transition angle between the day and nightside of a planet with the radius and surface gravity of HD\,209458\,b. The colors correspond to the temperature range (Blue: 500-1000\,K; Black: 1000-1500\,K; Red: 1000-1800\,K). See \fig{retrievalGJ} for other details. }
 \label{retrievalHD}
\end{figure}

The retrieved temperatures and water abundances as a function of the transition angle ($\openAngle$) for our GJ\,1214\,b case are shown in \fig{retrievalGJ}. \fig{retrievalHD} shows the temperature result for HD\,209458\,b. We tested different day and night side temperatures to see how planets with various irradiations would behave. To be able to compare the retrieved temperature ($\Tret$) in these different cases, we use the relative retrieved temperature 
\balign{
\Trel\equiv\frac{\Tret-\Tnight}{\Tday-\Tnight},}
which should thus be equal to 0.5 if we were to retrieve the terminator temperature. Although there are some quantitative differences among these cases, some robust trends emerge:
\begin{itemize}
    \item[$\bullet$] The retrieved temperature is systematically biased toward a higher temperature than that of the terminator ($\Trel\ge 0.5$). 
    \item[$\bullet$] There are two regimes separated by a critical day-night transition angle that depends on the characteristics of the planet and is roughly consistent with our estimate of the opening angle of the part of the planet that is probed in transit, i.e. the limb ($\limbAngle$, denoted by a vertical dashed line; See \fig{Cells}). 
    \item[$\bullet$] For $\openAngle< \limbAngle$, the retrieved temperature decreases roughly linearly with the transition angle. As the transition angle between the day and night side goes to zero (very sharp transition expected for the hottest planets), the retrieved temperature approaches the dayside temperature. 
    \item[$\bullet$] For $\openAngle>\limbAngle$, the temperature structure within the limb, hence the retrieved temperature, does not vary much. Whether the actual retrieved temperature is equal to the temperature at the terminator depends on the case (see below).
    \item[$\bullet$] Despite the uniform composition in our models, the retrieved abundance is always significantly biased -- in the sense that the real abundance is outside the formal error bars of the retrieval -- although the magnitude and direction depends on the specific case. It is sensible to assume that other more complex biases will arise if chemical gradients are present as well.
\end{itemize}

If in the HD\,209458\,b case (see \fig{retrievalHD}), the retrieved temperature converges toward the temperature at the terminator when the latter becomes more uniform ($\openAngle \rightarrow 180^\circ$), it is not necessarily the case for GJ\,1214\,b. We find that this absence of convergence at large angles always occurs when the hot, deep atmosphere below the $\piso$ level is probed by the transit spectrum: the retrieval is biased by the \textit{vertical} temperature gradient. This does not happen for our Hot Jupiter case because of the larger radius that push the transit photosphere at lower pressures. Although an important bias in itself, it has already been studied by \citet{RWV16}, and will not be further discussed here. 
%\JL{I think I know why the radius is overestimated at large opening angle: the retrieval sees the temperature of the upper limb which is smaller than the temperature below. However, it computes the 10 bar radius needed to get the observed effective radius by integrating the atmosphere downward (matter of speaking) assuming the retrieved temperature, hence a lower scale height. It thus needs to have a model bottom that is at a higher R.}

%\JL{problem, this has nothing to do with our 3D effect and would happen with a 1D isothermal retrieval over a non isothermal atmosphere. I would thus suggest to get rid of the radius plot which is just confusing. What do you think?}

\subsection{Could we see that something is wrong?}

\begin{figure}
   \centering
   \includegraphics[width=9cm]{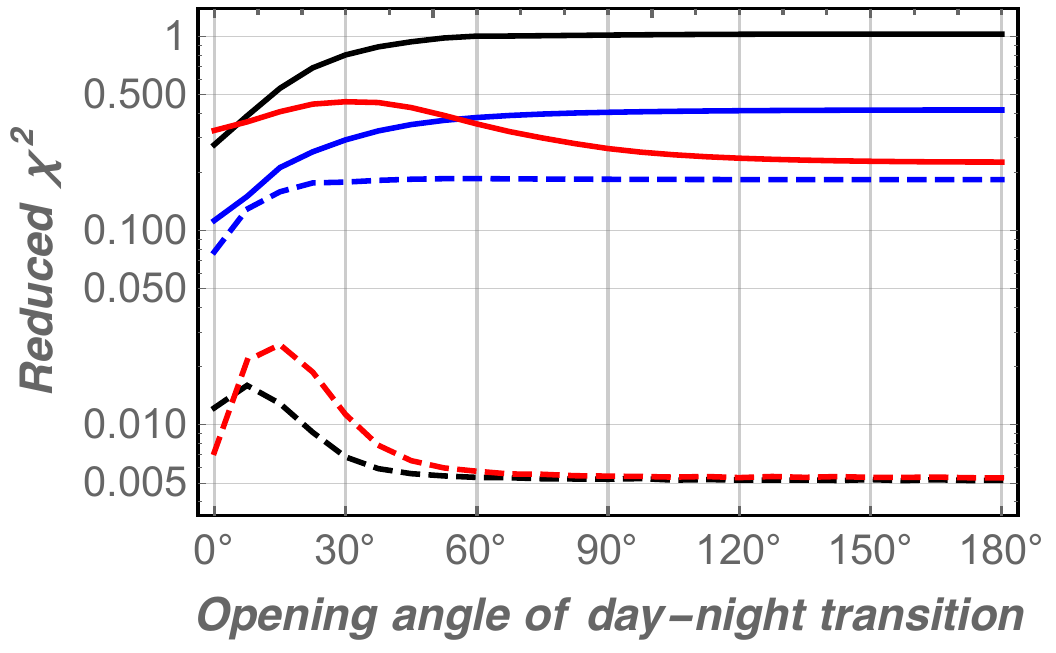}
      \caption{Reduced $\chi^2$ for the fit of the optimal retrieved models from \figs{retrievalGJ}{retrievalHD}. The color coding is the same as for those figures. Solid curves are for GJ\,1214\,b and dashed are for HD\,209458\,b. As the reduced $\chi^2$ is always close to or smaller than 1, the fit would always be considered satisfactory. }
         \label{chi2}
   \end{figure}
   
Could an observer, having performed the retrieval, detect that the retrieved quantities are biased by the day to night temperature gradient? This is indeed a crucial point.

Unfortunately, this seems precluded, even with the exquisite precision of JWST. As can be seen in \fig{fig:posterior}, the best-fit 1D, isothermal spectrum does not miss any feature of the input 3D spectrum. In fact, \fig{chi2} shows that the reduced $\chi^2$ of the optimal retrieved models are always near or below unity\footnote{Finding reduced $\chi^2$ smaller than unity is allowed by the fact that, although the noise-induced uncertainty on the input spectrum is used as an input of the retrieval procedure, no noise instance has been added to the input spectrum itself (see \sect{sec:bias}). Contrary to the case of a comparison with real data, low $\chi^2$ here are not a sign of over-fitting.}. Based on this metric, the 1D isothermal atmosphere model thus provides an acceptable fit to the data, at least in the low resolution mode that we have explored here. In fact, it even seems that the fit is better when the bias is the strongest (low day-night transition angle). This counter intuitive result comes from the fact that when the temperature transition is sharp, we probe almost exclusively the dayside. The atmosphere thus appears more homogeneous.

A procedure that would use a radiative transfer code similar to Pythmosph3R to retrieve a 3D structure/composition (which would imply formidable computing resources) would admittedly reveal the issue as its posterior distributions would expose the full extent of the degeneracies and result in larger, more reliable error bars on the retrieved quantities. Nevertheless, such a sophisticated tool may not be able to achieve a better retrieval. Indeed, even if a better $\chi^2$ may be found with a forward model using a 3D thermal and compositional structure, the 1D model already provides an excellent match and the improvement, if any, would be achieved at the expense of adding so many parameters that parsimony criteria may favour the most simple model.

However, a much higher resolution may change this state of affairs. Especially, if we start to be sensitive to the line shape of individual lines. It is also possible that an east-west asymmetry of the terminator could add some signal that would be distinguishable from any 1D profile.
This will have to be assessed in a future study.

\section{Which atmospheres are affected?}

Since the first detections of the thermal emission of a planet, the existence of a strong day-night temperature gradient on hot extrasolar planets has been well established \citep{CA11}. A clear trend has even emerged that the hotter the planet, the greater the thermal contrast \citep{KST17,KC18} --- a thermal contrast that can reach more than a thousand degrees. The bias on the retrieved limb temperatures on real planets is thus potentially huge. The yet rather observationally unconstrained parameter is the width of the day-night transition region (here $\openAngle$) and how it compares to the width of the limb that is effectively probed in transmission ($\limbAngle$).
Since phase curve observations have not quite yet the resolution necessary to precisely measure the width of the transition, we turn to published models. 

For Neptune-like planets such as GJ\,1214\,b, our predictions based on a GCM tell us that the thermal-only effect of the day-night temperature gradient is on the order of 50-100 ppm in the water bands, and clearly detectable by JWST (see \fig{Spec}). It is thus sensible to assume that any hotter Neptune-like planet should be increasingly affected because of their higher $\H/\Rp$ ratio and their stronger thermal day-night contrast (see \fig{spectrumdiff}). On the contrary, how cold -- the stratosphere of colder planets being more uniform -- will a planet need to be before such effects are undetectable remains to be elucidated.

However, keep in mind that we discussed here only the direct thermal effect, leaving out the possible chemical heterogeneities. The strong signature of day-night gradients in the CO$_2$ and CH$_4$ bands the spectrum in \fig{Spec}, which can reach 600-1000\,ppm, are believed to be of even greater importance and could possibly strongly hamper the ability of conventional retrieval algorithm to retrieve meaningful molecular abundances in the case of heterogeneous atmospheres.

Such considerations are even more important for hot-Jupiters. Despite their higher gravity, they can be much hotter, which entails that the day night contrasts are expected to be much stronger---both thermally and compositionally because chemical timescales are expected to be short compared to the advection timescales. A good example are the so-called ultra hot-Jupiters where it is predicted that some very abundant molecules on the nightside, such as water, could be almost completely absent on the dayside due to thermal dissociation \citep{PLB18}.

\section{Conclusion}

Overall, our most important conclusion is that the region of the atmosphere probed in by transit spectroscopy, i.e. the limb, is not confined to a narrow annulus around the planet as often implied, but can indeed extend relatively far throughout its two hemispheres. This is especially true for hot and/or low gravity objects, the most significant metric being the ratio of the atmospheric scale height to the radius of the planet. 

As a result, in addition to the variations of atmospheric properties of the atmosphere \textit{along} the terminator, the transit spectrum is also affected by their variations \textit{across} the limb, i.e. along the path of the light rays.

To investigate all these effects, we have developed Pytmosph3R, a transit spectrum generator that can work with a 3D atmospheric structure, whether it is the output of a global circulation model or a more idealized one.
Using this tool along with a 3D atmospheric model of GJ\,1214\,b, we have recovered previous results that the temperature and compositional variations along the terminator significantly affect the transit spectrum and will have to be accounted for in future studies. These effects can in principle be partially accounted for by using a (1+1)D, or limb integrated, approach where one 1D spectrum is generated for each part of the limb before it is weighted and added to the others to generate the global one. 

However, our fully 3D framework has shown that at the precision that will be reached by future observatories, the limb integrated approach is insufficient. Indeed, we have shown that for temperature gradients realistically expected for observable exoplanets, the transit spectrum is significantly affected by the structure of the atmosphere across the limb, i.e. the thermal and compositional gradients between the day and night side of the planet. 
We further demonstrated that this effect systematically biases 1D retrieval methods toward the temperature of the day side. The extent of this bias, however, depends on the strength of the temperature contrast as well as its sharpness around the terminator, the latter being the most difficult to predict.

In other words, one should be aware of the fact that the temperature (or its profile) retrieved from transmission spectra may not apply to the terminator itself, and that temperatures at the terminator are in fact significantly smaller. This will of course be a routine problem for future high precision observatories, but we have demonstrated that the effect on the spectrum -- which can reach hundreds, if not thousands of ppms in some cases (see \fig{spectrumdiff}) -- is well above the precision of 50-100\,ppm that has been achieved with HST and Spitzer \citep{Cowan2015}. So current observations of some Hot Jupiter are possibly already affected. To what extent? this remains to be elucidated.

\begin{acknowledgements}
 This project has received funding from the European Research Council (ERC) under the European Union's Horizon 2020 research and innovation programme (grant agreement n$^\circ$ 679030/WHIPLASH, n$^\circ$ 758892/ExoAI, n$^\circ$ 776403/ExoplANETS A  and under the European Union's Seventh Framework Programme (FP7/2007-2013)/ ERC grant agreement n$^\circ$ 617119/ExoLights. IPW furthermore acknowledge funding by the Science and Technology Funding Council (STFC) grants: ST/K502406/1 and ST/P000282/1.) 
\end{acknowledgements}

\appendix

\section{Slant optical depth and transit radius of an isothermal, gray atmosphere with constant gravity}\label{app:guillotmodel}

\subsection{Homogeneous case}

We first start be computing the optical depth along a ray crossing the atmosphere of a planet where temperature, gravity, and composition are constant. Here, we will follow the notations of \citet{VCM14}, but the reader is referred to \citet{Guigui} for a more in depth discussion. Notations are summarized in \fig{schematic_transit}.

The constant atmospheric scale height, $\H$, entails that the gas number density is given by \balign{\ngas=\ngasref e^{-\z/\H},}
where $\ngasref$ is the density at the reference radius ($\Rp$). For a ray with a given tangent altitude $\zt$, the altitude in the atmosphere at a distance $\x$ from the limb plane is given by 
\balign{\z=\zt+ \frac{\x^2}{2\Rp},}
to first order in $\zt/\Rp$. The optical depth along a ray from the star to a given position $\x$ due to a given species is thus 
\balign{
\tautr(\zt)&=\int_{-\infty}^{\x}  \sigmol \,\chi\, \ngasref\,e^{-\zt/\H} e^{-\x'^2/(2 \Rp \H)} \d \x'\nonumber\\
&=\sigmol\,\chi \,\ngasref\, e^{-\zt/\H} \int_{-\infty}^{\x}   e^{-\x'^2/(2 \Rp \H)} \d \x'\nonumber\\
&=\sqrt{2\pi \Rp \H}\,\sigmol\,\chi \,\ngasref\, e^{-\zt/\H} \left(\frac{1}{2}+\frac{1}{2}\mathrm{erf}\left(\frac{\x}{\sqrt{2 \Rp \H}}\right)\right) \nonumber\\
&\xrightarrow{\x\rightarrow\infty}\sqrt{2\pi \Rp \H}\,\sigmol\,\chi \,\ngasref\, e^{-\zt/\H} , \label{tautransit}
}
where $\chi$ is the volume mixing ration of the considered species, and $\sigmol$ its cross section at the wavelength considered. Noting that the vertical optical depth is given by
\balign{\tauvert(\zt)=\H\,\sigmol\,\chi \,\ngasref\, e^{-\zt/\H},}
we retrieve the result from \citet{Guigui} that
\balign{\tautr= \left(\frac{2\pi \Rp}{\H}\right)^{\frac{1}{2}}\tauvert.}
Following \sect{observables}, to first order, the transit depth is given by 
\balign{
\delta=\Rs^{-2}\left(\Rp^2+2\int_{\Rp}^{\infty} \left(1-e^{-\tautr(\rho)}\right)\rho \d\rho \right), \label{analyticaltransit}
}
where $\rho=\Rp+\zt$. Along with \eq{tautransit}, \eq{analyticaltransit} is used to validate our model in \sect{sec:validationguillot}. 

\subsection{Heterogeneous composition}

In this subsection, we will slightly modify the model above to answer the following question: how far from the limb plane can an increase in the abundance of a given species still affect the transit in the relevant bands. Two answer that, we will assume that the mixing ratio of the considered species is $\chid$ along the line ray for $\x < \xlimb$ (where $\xlimb$ is negative if the transition is on the day side) and $\chin$ beyond that. This is supposed to mimic a situation where an absorber, like TiO, becomes less abundant at the terminator and on the night side because it condenses at cooler temperatures. Because of the symmetry of the problem, this also treats the situation where an absorber becomes more abundant on the night side. Following \fig{schematic_transit}, $\xlimb$ is also parametrized by the limb angle $\tan (\limbAngle/2)= \xlimb / (\Rp+\zt)$.

If the composition were uniform, the optical depth at a given tangent altitude would be
\balign{\tautr=\sqrt{2\pi \Rp \H}\,\sigmol\,\chin \,\ngasref\, e^{-\zt/\H}.}
Using $\tautr\sim 1$ as a criterion for the effective altitude of absorption of the atmosphere, we get an effective altitude for the uniform case that is 
\balign{\z_\mathrm{uni}=\H \ln\left(\sqrt{2\pi \Rp \H}\,\sigmol\,\chin \,\ngasref \right).}

Now, in the heterogeneous case, using the penultimate line of \eq{tautransit} yields
\balign{\tautr&=\sqrt{2\pi \Rp \H}\,\sigmol \,\ngasref\, e^{-\zt/\H} \times \nonumber \\
&\hspace{1cm}\times\left(\frac{\chid+\chin}{2}+\frac{\chid-\chin}{2}\mathrm{erf}\left(\frac{\xlimb}{\sqrt{2 \Rp \H}}\right)\right),}
which is equal to the uniform case in the $\xlimb \rightarrow -\infty$ limit, as expected. The effective altitude is then
\balign{\z_\mathrm{het}&=\z_\mathrm{uni}-\H\ln 2 \nonumber\\
&+\H \ln\left(\left(\frac{\chid}{\chin}+1\right)+\left(\frac{\chid}{\chin}-1\right)\mathrm{erf}\left(\frac{\xlimb}{\sqrt{2 \Rp \H}}\right)\right) .}

At what angular distance from the limb plane can we still see such an increase of a given species in the transit spectrum? To answer this question, one needs to quantify at which angle $\limbAngle$ does the resulting change in effective transit altitude of the atmosphere due to the heterogeneity become measurable. This writes
\balign{
\delta_\mathrm{het}-\delta_\mathrm{uni}>\sigobs \ \ \Leftrightarrow \ \  \z_\mathrm{het}(\limbAngle)-\z_\mathrm{uni}>\frac{\Rs^2}{2 \Rp} \sigobs, }
where $\sigobs$ is the relative precision level of the observations. Using the parameters for HD\,209458\,b and assuming a noise floor of 10\,ppm -- which is probably conservative for JWST -- we see that an increase of the TiO abundance on the dayside of only a factor 100 is visible as far as 15$^\circ$ from the limb plane. This increase is also conservative as the abundance of TiO at temperature below 1600\,K is less that $10^{-10}$ \citep{Lod02}. This results in a limb width $\sim 30^\circ $ which is consistent with our other estimate (see \fig{Cells}).

\section{Finding the spherical coordinates of a cell in the cylindrical grid}\label{app:coordsystem}

To link our two coordinate systems, we will use a cartesian reference frame centered around the center of the planet and whose orthonormal reference axes are $\{\vX,\vY,\vZ\}$. $\vZ$ is the unit vector along the rotation axis of the planet (pointing toward the north pole). $\vX$ points toward a reference point at the equator which will be the origin of longitudes. $\vY$ is chosen to have a direct basis. The coordinates of any point in this system are $\vR=(\X,\Y,\Z)$.

Given the position of the observer 
\balign{
\vx \equiv\left(\begin{array}{c}\Xobs \\ \Yobs \\ \Zobs \end{array}\right)_{\vX,\vY,\vZ} =  \left(\begin{array}{c} \sin \colatobs \cos \lonobs \\ \sin \colatobs \sin \lonobs \\ \cos \colatobs \end{array}\right) ,\label{defobs} 
}
we need to know what are the physical conditions in the atmosphere at any given point $\vR$ determined by its cylindrical coordinates $(\rc,\th,\x)$. For this we need to find the set of spherical coordinates $(\rs,\lon,\colat)$ corresponding to $\vR$.

This can be done by first noticing that with our definitions (see \sect{orientation}), $\vR$ can be decomposed into a component in the plane of the sky and one along the line of sight
\balign{\vR(\rc,\th,\x)=\vray(\rc,\th) + \x\, \vx, \label{defray}}
where $\vray=(\Xray,\Yray, \Zray)$ is the intersection between a $(\rc,\th)$-ray and the plane of the sky.

The first step is to compute $\Xray,\,\Yray$, and $\Zray$. These are uniquely determined thanks to the three following definitions that can be combined into one degree-two equation:
\begin{itemize}
    \item[$\bullet$] Since $\th$ is the angle between the projection of the planetary rotation vector onto the plane of the sky and $\vray$, it can be shown that $\Zray =  \rc\,\sin\colatobs \,\cos \th.$ 
    \item[$\bullet$] $\vray$ is in the plane of the sky so that $\vray\cdot \vx=0=\Xray \Xobs +\Yray \Yobs+ \Zray\Zobs $.

    \item[$\bullet$] By definition, $\rc\equiv | \vray | = \sqrt{\Xray ^2+\Yray ^2+ \Zray ^2}$.
\end{itemize}

Once these three components are known for each $(\rc,\th)$, the spherical coordinates of a point are given by solving
\balign{
\rs^2&=\rc^2+\x^2,\label{radiusequation} \\
\lon&= \arctan\left( \frac{\Yray(\rc,\th)+ \x \Yobs}{\Xray(\rc,\th)+ \x \Xobs}\right), \label{longitudeequation}\\
\colat&= \arccos\left( \frac{\Zray(\rc,\th)+ \x \Zobs}{\rs}\right).\label{latitudeequation}
}

Finally, for numerical reasons, we determine the set of indices $(i_\rs,i_\lon,i_\lat)$ representing this specific cell in the spherical grid.

To give a concrete example, for the simple case of a synchronous planet for which the origin of longitudes is chosen at the substellar point and observed when the star, planet, and observer are perfectly aligned, we have $\vx =\left(-1,0,0\right).$ Then, solving the equations above yields $\vray(\rc,\th)=(0,\rc \sin \th, \rc \cos \th),$ and the correspondance relationship is
\balign{
\rs^2&=\rc^2+\x^2,\\
\lon&= \arctan\left( \frac{\rc \sin \th}{- \x }\right),\\
\colat&= \arccos\left( \frac{\rc \cos \th}{\rs}\right).
}

\section{Computing the position of the intersection of a ray with a given interface of the spherical grid}
\label{app:intersections}

In a spherical grid, the separation between cells is done by three types of surfaces: spheres, planes of constant longitude (meridians), and cones of constant (co)latitude. Our goal is to compute the location along a ray ($\x_\mathrm{int}$) of the intersection of this ray with any given of those surfaces. Because we know the indices of the cells before and after the intersection, we always know what type of surface we are crossing, and the value of the constant radius/longitude/colatitude identifying this surface (respectively $\rs_\mathrm{int}$, $\lon_\mathrm{int}$, and $\colat_\mathrm{int}$).

Because we also know the ($\rc,\th$)-ray we are dealing with and the observer's location, bear in mind that both $\vray=(\Xray,\Yray, \Zray)$ and $\vx=(\Xobs,\Yobs, \Zobs)$ are known.

The equations to be solved are
\begin{itemize}
    \item[$\bullet$] Intersection with a sphere:
    \balign{\x_\mathrm{int}=\pm\sqrt{\rs_\mathrm{int}^2-\rc^2}.}
    The value is positive between the limb plane and the observer and negative otherwise.
    \item[$\bullet$] Intersection with a meridian: Using \eq{longitudeequation}, we can solve for the intersection, which yields
    \balign{\x_\mathrm{int}= \frac{\Yray(\rc,\th)-  \Xray(\rc,\th)\tan \lon_\mathrm{int}}{   \Xobs\tan \lon_\mathrm{int}-\Yobs}.}
    
    \item[$\bullet$]Intersection with a cone of constant colatitude: Combining \eqs{radiusequation}{latitudeequation} we get a second degree equation in $\x_\mathrm{int}/\rc$ 
    \balign{
0&=\left(\cos^2\colat_\mathrm{int}-\cos^2\colatobs\right) \,\left(\frac{\x_\mathrm{int}}{\rc}\right)^2  \nonumber\\
&- \left(2\cos\colatobs \sin\colatobs \cos\th \right) \,\left(\frac{\x_\mathrm{int}}{\rc}\right) \nonumber\\&
+\left( \cos^2\colat_\mathrm{int} -\sin^2\colatobs \cos^2\th \right).  
}
As can be seen from the equation above, the equation is the same for $\colat_\mathrm{int}$ and $\pi-\colat_\mathrm{int}$ (i.e. $\pm \lat_\mathrm{int}$). This means that one cannot know \textit{a priori} whether the solutions found are in the northern or southern hemisphere. In fact, when two solutions exist, the ray either intersects the cone twice in the same hemisphere (when $|\latobs|<|\lat_\mathrm{int}|$), or once on each side of the equator. To remove this degeneracy, one can compute the position of the intersection of the ray with the equator $\x_\mathrm{equ}$. Then if $\x_\mathrm{int}>\x_\mathrm{equ}$ the intersection is in the same hemisphere than the observer and vice versa. 
\end{itemize}

\section{Rayleigh scattering data}\label{rayleighscatdata}

Typically, refractive indices follow the generic expression 
\balign{(n-1)10^8=A+\frac{B}{C-\lambda^{-2}},}
with terms and their values as described in \tab{index}, along with the corresponding King correction factor equations. This term is taken as unity for mono-atomic gases and is calculated $ab$ $initio$ as described in \citet{Bates} for diatomic gases. The wavelength dependency of variables is also specified in \tab{index}. Note that for all the formulae in this section, the wavelength is expressed in $\mu$m. 

Some molecules have non-standard parametrizations. 
For $CO_2$ (\citet{Sneep})
\balign{
&\left(\frac{n_\lambda^2-1}{n_\lambda^2+2}\right)\times 10^8=1.1427\times10^{14}\times \nonumber\\
&\times(\frac{5.79925\times10^{-5}}{5.0821\times10^{2}-1/\lambda^2}+\frac{1.2005\times10^{-6}}{7.9608\times10^1-1/\lambda^2} + \nonumber\\
&+\frac{5.3334\times10^{-8}}{5.6306\times10^1-1/\lambda^2} +\frac{4.3244\times10^{-8}}{4.619\times10^1-1/\lambda^2}+\nonumber\\
&+\frac{1.2181\times10^{-13}}{5.8474\times10^{-2}-1/\lambda^2})
}
and
\balign{F_k=1.1364+\frac{2.53\times10^{-3}}{\lambda^2}}

For CH$_4$ (\citet{Sneep})
\balign{(n_\lambda-1)=4.6662.10^{-4}+\frac{4.02\times10^{-6}}{\lambda^2}}

For H$_2$O, if $\lambda>0.23 \mu m$
\balign{(n_\lambda-1)=\frac{4.92303\times10^{-2}}{2.380185.10^{2}-1/\lambda^2}+\frac{1.42723\times10^{-3}}{5.73262\times10^{1}-1/\lambda^2}}
and if $\lambda<0.23 \mu m$
\balign{
(n_\lambda-1)&=6.85143\times10^{-2}+ \nonumber\\
&+\frac{2.10884\times10^{-2}}{1.32274\times10^{2}-1/\lambda^2}+\frac{1.4837\times10^{-3}}{3.932957\times10^{1}-1/\lambda^2} 
}

For molecular hydrogen we calculate the cross-section as described in \citet{DW62}, if $\lambda>30 \mu m$
\balign{\sigma_s(\nu)=\frac{8.49\times10^{-33}}{\lambda^4}}
and if $\lambda<30 \mu m$
\balign{
\sigma_s(\nu)=\frac{8.14\times10^{-33}}{\lambda^4}+\frac{1.28\times10^{-34}}{\lambda^6}+\frac{1.61\times10^{-35}}{\lambda^8}
}

%%%%%%%%%%%%%%%%%%%%%%%%%%%%%%%%%%%%%%%%
%%%% Table 1
%%%%%%%%%%%%%%%%%%%%%%%%%%%%%%%%%%%%%%%% 
\begin{table*}
\caption[]{Values of the parameters used in Eq.(6).}
\centering
\label{index}
\begin{tabular}{|c c c c c c l| }     % 7 columns 
\hline\hline       
Gas & Wavelength ($\mu m$) & $A/10^4$ & $B/10^6 (\mu \mathrm{m}^{-2})$ & $C/10^2(\mu \mathrm{m}^{-2})$ & King factor $F_k$ & References\\ 
\hline                    
   He & $all$ & $0.2283$ & $0.2283$ & $1.532$ & $1$ & \citet{Thalm}\\  
   N$_2$ & $\lambda>0.460$ & $0.6498$    & $3.0740$ & $1.44$ & & \citet{Thalm}\\
    & $0.460>\lambda>0.254$ & $0.6677$    & $3.1882$ & $1.44$ &  $1.034+3.17\times10^{-4}/\lambda^2$ &\citet{Bates} \\
    & $\lambda<0.254$ & $0.6999$    & $3.2336$ & $1.44$ & & \\
   O$_2$ & $\lambda>0.546$ & $2.1351$     & $0.218567$ & $0.409$ & & \citet{Thalm}\\
    & $0.546<\lambda<0.288$ & $2.0564$     & $0.248090$ & $0.409$ & $1,096+1.385\times10^{-3}/\lambda^2$& \citet{Bates}\\
    & $0.288>\lambda<0.221$ & $2.21204$     & $0.203187$ & $0.409$ & $+1.448\times10^{-4}1/\lambda^4$ & Sneep, M. (2004)\\
    & $\lambda<0.221$ & $2.37967$     & $0.268988$ & $0.409$ & & \\
   CO & $all$ & $2.2851$    & $0.0456$ & $0.51018$ & $1.016$ & Sneep, M. (2004) \\
   Ar & $all$ & $0.6432135$    & $0.028606$ & $1.44$ & $1$ & \citet{Thalm}\\
\hline\hline                 
\end{tabular}
\end{table*}

\bibliography{references} % your references Yourfile.bib
%\bibliography{biblio} % your references Yourfile.bib
\bibliographystyle{aa} % style aa.bst

\end{document}